# LSMR: AN ITERATIVE ALGORITHM FOR SPARSE LEAST-SQUARES PROBLEMS*

DAVID CHIN-LUNG FONG† AND MICHAEL SAUNDERS‡

**Abstract.** An iterative method LSMR is presented for solving linear systems $Ax = b$ and least-squares problems $\min \|Ax-b\|_2$, with $A$ being sparse or a fast linear operator. LSMR is based on the Golub-Kahan bidiagonalization process. It is analytically equivalent to the MINRES method applied to the normal equation $A^TAx = A^Tb$, so that the quantities $\|A^Tr_k\|$ are monotonically decreasing (where $r_k = b - Ax_k$ is the residual for the current iterate $x_k$). We observe in practice that $\|r_k\|$ also decreases monotonically, so that compared to LSQR (for which only $\|r_k\|$ is monotonic) it is safer to terminate LSMR early. We also report some experiments with reorthogonalization.

## 1. Introduction.

We present a numerical method called LSMR for computing a solution $x$ to the following problems:

$$\begin{array}{ll} \text{Unsymmetric equations:} & \text{minimize } \|x\|_2 \text{ subject to } Ax = b \\ \text{Linear least squares (LS):} & \text{minimize } \|Ax-b\|_2 \\ \text{Regularized least squares:} & \text{minimize } \left\| \begin{pmatrix} A \\ \lambda I \end{pmatrix} x - \begin{pmatrix} b \\ 0 \end{pmatrix} \right\|_2 \end{array}$$

where $A \in \mathbb{R}^{m\times n}$, $b \in \mathbb{R}^m$, and $\lambda \ge 0$, with $m \le n$ or $m \ge n$. The matrix $A$ is used as an operator for which products of the form $Av$ and $A^Tu$ can be computed for various $v$ and $u$. (If $A$ is symmetric or Hermitian and $\lambda = 0$, MINRES-QLP [4] is applicable.)

LSMR is similar in style to the well known method LSQR [16, 17] in being based on the Golub-Kahan bidiagonalization of $A$ [6]. LSQR is equivalent to the conjugate-gradient (CG) method applied to the normal equation $(A^TA+\lambda^2I)x = A^Tb$. It has the property of reducing $\|r_k\|$ monotonically, where $r_k = b - Ax_k$ is the residual for the approximate solution $x_k$. (For simplicity, we are letting $\lambda = 0$.) In contrast, LSMR is equivalent to MINRES [15] applied to the normal equation, so that the quantities $\|A^Tr_k\|$ are monotonically decreasing. In practice we observe that $\|r_k\|$ also decreases monotonically, and is never very far behind the corresponding value for LSQR. Hence, although LSQR and LSMR ultimately converge to similar points, it is safer to use LSMR in situations where the solver must be terminated early.

Stopping conditions are typically based on *backward error*: the norm of some perturbation to $A$ for which the current iterate $x_k$ solves the perturbed problem exactly. Experiments on many sparse LS test problems show that for LSMR, a certain *cheaply computable* backward error for each $x_k$ is close to the *optimal* (smallest possible) backward error. This is an unexpected but highly desirable advantage.

†ICME, Stanford University (clfong@stanford.edu). Partially supported by a Stanford Graduate Fellowship.

‡Systems Optimization Laboratory, Department of Management Science and Engineering, Stanford University, CA 94305-4026 (saunders@stanford.edu). Partially supported by Office of Naval Research grant N00014-08-1-0191 and by the U.S. Army Research Laboratory, through the Army High Performance Computing Research Center, Cooperative Agreement W911NF-07-0027.

**1.1. Overview.** Section 2 introduces the Golub-Kahan process and derives the basic LSMR algorithm with $\lambda = 0$. Section 3 derives various norms and stopping criteria. Section 4 discusses singular systems and complexity. Section 5 derives the LSMR algorithm with $\lambda \geq 0$. Section 6 describes backward error estimates. Section 7 gives numerical results on a range of overdetermined and square systems. Section 8 summarizes our findings, and Appendix A proves one of the main lemmas.

**1.2. Notation.** Matrices are denoted by $A, B, \ldots,$ vectors by $v, w, \ldots,$ and scalars by $\alpha, \beta, \ldots$. Two exceptions are $c$ and $s$, which denote the significant components of a plane rotation matrix, with $c^2 + s^2 = 1$. For a vector $v$, $\|v\|$ always denotes the 2-norm of $v$. For a matrix $A$, $\|A\|$ usually denotes the Frobenius norm, and the condition number of a matrix $A$ is defined by $\mathrm{cond}(A) = \|A\|\|A^+\|$, where $A^+$ denotes the pseudoinverse of $A$. Vectors $e_1$ and $e_k$ denote columns of an identity matrix. Items like $\dot{\beta}_k$ and $\ddot{\beta}_k$ are about to change to something similar like $\tilde{\beta}_k$.

## 2. Derivation of LSMR.

We begin with the Golub-Kahan process [6], an iterative procedure for transforming $\begin{pmatrix} b & A \end{pmatrix}$ to upper-bidiagonal form $\begin{pmatrix} \beta_1 e_1 & B_k \end{pmatrix}$.

**2.1. The Golub-Kahan process.**

1. Set $\beta_1 u_1 = b$ (shorthand for $\beta_1 = \|b\|$, $u_1 = b/\beta_1$) and $\alpha_1 v_1 = A^T u_1$.
2. For $k = 1, 2, \ldots,$ set

$$\beta_{k+1} u_{k+1} = A v_k - \alpha_k u_k \quad \text{and} \quad \alpha_{k+1} v_{k+1} = A^T u_{k+1} - \beta_{k+1} v_k. \tag{2.1}$$

After $k$ steps, we have

$$AV_k = U_{k+1} B_k \quad \text{and} \quad A^T U_{k+1} = V_{k+1} L_{k+1}^T,$$

where we define $V_k = \begin{pmatrix} v_1 & v_2 & \ldots & v_k \end{pmatrix}$, $U_k = \begin{pmatrix} u_1 & u_2 & \ldots & u_k \end{pmatrix}$, and

$$B_k = \begin{pmatrix} \alpha_1 & & & \\ \beta_2 & \alpha_2 & & \\ & \ddots & \ddots & \\ & & \beta_k & \alpha_k \\ & & & \beta_{k+1} \end{pmatrix}, \qquad L_{k+1} = \begin{pmatrix} B_k & \alpha_{k+1} e_{k+1} \end{pmatrix}.$$

Now consider

$$\begin{aligned} A^T A V_k = A^T U_{k+1} B_k = V_{k+1} L_{k+1}^T B_k &= V_{k+1} \begin{pmatrix} B_k^T \\ \alpha_{k+1} e_{k+1}^T \end{pmatrix} B_k \\ &= V_{k+1} \begin{pmatrix} B_k^T B_k \\ \alpha_{k+1} \beta_{k+1} e_k^T \end{pmatrix}. \end{aligned}$$

This is equivalent to what would be generated by the symmetric Lanczos process with matrix $A^T A$ and starting vector $A^T b$. (For this reason we define $\bar{\beta}_k \equiv \alpha_k \beta_k$ below.)

**2.2. Using Golub-Kahan to solve the normal equation.** Krylov subspace methods for solving linear equations form solution estimates $x_k = V_k y_k$ for some $y_k$, where the columns of $V_k$ are an expanding set of theoretically independent vectors. (In this case, $V_k$ and also $U_k$ are theoretically orthonormal.)

For the equation $A^T A x = A^T b$, any solution $x$ has the property of minimizing $\|r\|$, where $r = b - Ax$ is the corresponding residual vector. Thus, in the development of LSQR it was natural to choose $y_k$ to minimize $\|r_k\|$ at each stage. Since

$$r_k = b - AV_k y_k = \beta_1 u_1 - U_{k+1} B_k y_k = U_{k+1}(\beta_1 e_1 - B_k y_k),$$

where $U_{k+1}$ is theoretically orthonormal, the subproblem $\min_{y_k} \|\beta_1 e_1 - B_k y_k\|$ easily arose. In contrast, for LSMR we wish to minimize $\|A^T r_k\|$. Let $\bar\beta_k \equiv \alpha_k \beta_k$ for all $k$. Since $A^T r_k = A^T b - A^T A x_k = \beta_1 \alpha_1 v_1 - A^T A V_k y_k$, we have

$$A^T r_k = \bar\beta_1 v_1 - V_{k+1} \begin{pmatrix} B_k^T B_k \\ \alpha_{k+1}\beta_{k+1} e_k^T \end{pmatrix} y_k = V_{k+1} \left( \bar\beta_1 e_1 - \begin{pmatrix} B_k^T B_k \\ \bar\beta_{k+1} e_k^T \end{pmatrix} y_k \right),$$

and we are led to the subproblem

$$\min_{y_k} \|A^T r_k\| = \min_{y_k} \left\| \bar\beta_1 e_1 - \begin{pmatrix} B_k^T B_k \\ \bar\beta_{k+1} e_k^T \end{pmatrix} y_k \right\|. \tag{2.2}$$

Efficient solution of this LS subproblem is the heart of algorithm LSMR.

**2.3. Two QR factorizations.** As in LSQR, we form the QR factorization

$$Q_{k+1} B_k = \begin{pmatrix} R_k \\ 0 \end{pmatrix}, \qquad R_k = \begin{pmatrix} \rho_1 & \theta_2 & & \\ & \rho_2 & \ddots & \\ & & \ddots & \theta_k \\ & & & \rho_k \end{pmatrix}. \tag{2.3}$$

If we define $t_k = R_k y_k$ and solve $R_k^T q_k = \bar\beta_{k+1} e_k$, we have $q_k = (\bar\beta_{k+1}/\rho_k) e_k = \varphi_k e_k$ with $\rho_k = (R_k)_{kk}$ and $\varphi_k \equiv \bar\beta_{k+1}/\rho_k$. Then we perform a second QR factorization

$$\bar Q_{k+1} \begin{pmatrix} R_k^T & \bar\beta_1 e_1 \\ \varphi_k e_k^T & 0 \end{pmatrix} = \begin{pmatrix} \bar R_k & z_k \\ 0 & \bar\zeta_{k+1} \end{pmatrix}, \qquad \bar R_k = \begin{pmatrix} \bar\rho_1 & \bar\theta_2 & & \\ & \bar\rho_2 & \ddots & \\ & & \ddots & \bar\theta_k \\ & & & \bar\rho_k \end{pmatrix}. \tag{2.4}$$

Combining what we have with (2.2) gives

$$\begin{aligned} \min_{y_k} \|A^T r_k\| = \min_{y_k} \left\| \bar\beta_1 e_1 - \begin{pmatrix} R_k^T R_k \\ q_k^T R_k \end{pmatrix} y_k \right\| &= \min_{t_k} \left\| \bar\beta_1 e_1 - \begin{pmatrix} R_k^T \\ \varphi_k e_k^T \end{pmatrix} t_k \right\| \\ &= \min_{t_k} \left\| \begin{pmatrix} z_k \\ \bar\zeta_{k+1} \end{pmatrix} - \begin{pmatrix} \bar R_k \\ 0 \end{pmatrix} t_k \right\|. \end{aligned} \tag{2.5}$$

The subproblem is solved by choosing $t_k$ from $\bar R_k t_k = z_k$.

**2.4. Recurrence for $x_k$.** Let $W_k$ and $\bar W_k$ be computed by forward substitution from $R_k^T W_k^T = V_k^T$ and $\bar R_k^T \bar W_k^T = W_k^T$. Then from $x_k = V_k y_k$, $R_k y_k = t_k$, and $\bar R_k t_k = z_k$, we have $x_0 \equiv 0$ and

$$x_k = W_k R_k y_k = W_k t_k = \bar W_k \bar R_k t_k = \bar W_k z_k = x_{k-1} + \zeta_k \bar w_k.$$

**2.5. Recurrence for $W_k$ and $\bar W_k$.** If we write

$$\begin{aligned} V_k &= \begin{pmatrix} v_1 & v_2 & \cdots & v_k \end{pmatrix}, & W_k &= \begin{pmatrix} w_1 & w_2 & \cdots & w_k \end{pmatrix}, \\ \bar W_k &= \begin{pmatrix} \bar w_1 & \bar w_2 & \cdots & \bar w_k \end{pmatrix}, & z_k &= \begin{pmatrix} \zeta_1 & \zeta_2 & \cdots & \zeta_k \end{pmatrix}^T, \end{aligned}$$

an important fact is that when $k$ increases to $k+1$, all quantities remain the same except for one additional term.

The first QR factorization proceeds as follows. At iteration $k$ we construct a plane rotation operating on rows $l$ and $l+1$:

$$P_l = \begin{pmatrix} I_{l-1} & & & \\ & c_l & s_l & \\ & -s_l & c_l & \\ & & & I_{k-l-1} \end{pmatrix}.$$

Now if $Q_{k+1} = P_k \dots P_2 P_1$, we have

$$Q_{k+1} B_{k+1} = Q_{k+1} \begin{pmatrix} B_k & \alpha_{k+1} e_{k+1} \\ & \beta_{k+2} \end{pmatrix} = \begin{pmatrix} R_k & \theta_{k+1} e_k \\ 0 & \bar{\alpha}_{k+1} \\ & \beta_{k+2} \end{pmatrix},$$

$$Q_{k+2} B_{k+1} = P_{k+1} \begin{pmatrix} R_k & \theta_{k+1} e_k \\ 0 & \bar{\alpha}_{k+1} \\ & \beta_{k+2} \end{pmatrix} = \begin{pmatrix} R_k & \theta_{k+1} e_k \\ 0 & \rho_{k+1} \\ 0 & 0 \end{pmatrix}$$

and we see that $\theta_{k+1} = s_k \alpha_{k+1} = (\beta_{k+1}/\rho_k)\alpha_{k+1} = \bar{\beta}_{k+1}/\rho_k = \varphi_k$. Therefore we can write $\theta_{k+1}$ instead of $\varphi_k$.

For the second QR factorization, if $\bar{Q}_{k+1} = \bar{P}_k \dots \bar{P}_2 \bar{P}_1$ we know that

$$\bar{Q}_{k+1} \begin{pmatrix} R_k^T \\ \theta_{k+1} e_k^T \end{pmatrix} = \begin{pmatrix} \bar{R}_k \\ 0 \end{pmatrix},$$

and so

$$\bar{Q}_{k+2} \begin{pmatrix} R_{k+1}^T \\ \theta_{k+2} e_{k+1}^T \end{pmatrix} = \bar{P}_{k+1} \begin{pmatrix} \bar{R}_k & \bar{\theta}_{k+1} e_k \\ & \bar{c}_k \rho_{k+1} \\ & \theta_{k+2} \end{pmatrix} = \begin{pmatrix} \bar{R}_k & \bar{\theta}_{k+1} e_k \\ & \bar{\rho}_{k+1} \\ & 0 \end{pmatrix}. \tag{2.6}$$

By considering the last row of the matrix equation $R_{k+1}^T W_{k+1}^T = V_{k+1}^T$ and the last row of $\bar{R}_{k+1}^T \bar{W}_{k+1}^T = W_{k+1}^T$ we obtain equations that define $w_{k+1}$ and $\bar{w}_{k+1}$:

$$\begin{aligned} \theta_{k+1} w_k^T + \rho_{k+1} w_{k+1}^T &= v_{k+1}^T, \\ \bar{\theta}_{k+1} \bar{w}_k^T + \bar{\rho}_{k+1} \bar{w}_{k+1}^T &= w_{k+1}^T. \end{aligned}$$

**2.6. The two rotations.** To summarize, the rotations $P_k$ and $\bar{P}_k$ have the following effects on our computation:

$$\begin{pmatrix} c_k & s_k \\ -s_k & c_k \end{pmatrix} \begin{pmatrix} \bar{\alpha}_k & \\ \beta_{k+1} & \alpha_{k+1} \end{pmatrix} = \begin{pmatrix} \rho_k & \theta_{k+1} \\ 0 & \bar{\alpha}_{k+1} \end{pmatrix}$$

$$\begin{pmatrix} \bar{c}_k & \bar{s}_k \\ -\bar{s}_k & \bar{c}_k \end{pmatrix} \begin{pmatrix} \bar{c}_{k-1}\rho_k & & \bar{\zeta}_k \\ \theta_{k+1} & \rho_{k+1} & \end{pmatrix} = \begin{pmatrix} \bar{\rho}_k & \bar{\theta}_{k+1} & \zeta_k \\ 0 & \bar{c}_k \rho_{k+1} & \bar{\zeta}_{k+1} \end{pmatrix}.$$

**2.7. Speeding up forward substitution.** The forward substitutions for computing $w$ and $\bar{w}$ can be made more efficient if we define $h_k = \rho_k w_k$ and $\bar{h}_k = \rho_k \bar{\rho}_k \bar{w}_k$. We then obtain the updates described in part 6 of the pseudo-code below.

**2.8. Algorithm LSMR.** The following summarizes the main steps of algorithm LSMR for solving $Ax \approx b$, excluding the norms and stopping rules developed later.

1. (Initialize)

$$\begin{array}{cccccc} \beta_1 u_1 = b & \alpha_1 v_1 = A^T u_1 & \bar{\alpha}_1 = \alpha_1 & \bar{\zeta}_1 = \alpha_1 \beta_1 & \rho_0 = 1 & \bar{\rho}_0 = 1 \\ \bar{c}_0 = 1 & \bar{s}_0 = 0 & h_1 = v_1 & \bar{h}_0 = 0 & x_0 = 0 & \end{array}$$

2. For $k = 1, 2, 3 \dots$, repeat steps 3–6.
3. (Continue the bidiagonalization)

$$\beta_{k+1} u_{k+1} = A v_k - \alpha_k u_k, \qquad \alpha_{k+1} v_{k+1} = A^T u_{k+1} - \beta_{k+1} v_k$$

4. (Construct and apply rotation $P_k$)

$$\rho_k = \left(\bar{\alpha}_k^2 + \beta_{k+1}^2\right)^{\frac{1}{2}} \qquad c_k = \bar{\alpha}_k / \rho_k \qquad s_k = \beta_{k+1} / \rho_k \tag{2.7}$$

$$\theta_{k+1} = s_k \alpha_{k+1} \qquad \bar{\alpha}_{k+1} = c_k \alpha_{k+1} \tag{2.8}$$

5. (Construct and apply rotation $\bar{P}_k$)

$$\bar{\theta}_k = \bar{s}_{k-1} \rho_k \qquad \bar{\rho}_k = \left((\bar{c}_{k-1} \rho_k)^2 + \theta_{k+1}^2\right)^{\frac{1}{2}}$$

$$\bar{c}_k = \bar{c}_{k-1} \rho_k / \bar{\rho}_k \qquad \bar{s}_k = \theta_{k+1} / \bar{\rho}_k \tag{2.9}$$

$$\zeta_k = \bar{c}_k \bar{\zeta}_k \qquad \bar{\zeta}_{k+1} = -\bar{s}_k \bar{\zeta}_k \tag{2.10}$$

6. (Update $h$, $\bar{h}$ $x$)

$$\begin{aligned} \bar{h}_k &= h_k - (\bar{\theta}_k \rho_k / (\rho_{k-1} \bar{\rho}_{k-1})) \bar{h}_{k-1} \\ x_k &= x_{k-1} + (\zeta_k / (\rho_k \bar{\rho}_k)) \bar{h}_k \\ h_{k+1} &= v_{k+1} - (\theta_{k+1} / \rho_k) h_k \end{aligned}$$

**3. Norms and stopping rules.** Here we derive $\|r_k\|$, $\|A^T r_k\|$, $\|x_k\|$ and estimates of $\|A\|$ and $\text{cond}(A)$ for use within stopping rules. All quantities require $O(1)$ computation at each iteration.

**3.1. Computing $\|r_k\|$.** We transform $\bar{R}_k^T$ to upper-bidiagonal form using a third QR factorization: $\widetilde{R}_k = \widetilde{Q}_k \bar{R}_k^T$ with $\widetilde{Q}_k = \widetilde{P}_{k-1} \dots \widetilde{P}_1$. This amounts to one additional rotation per iteration. Now let

$$\tilde{t}_k = \widetilde{Q}_k t_k, \qquad \tilde{b}_k = \begin{pmatrix} \widetilde{Q}_k & \\ & 1 \end{pmatrix} Q_{k+1} e_1 \beta_1. \tag{3.1}$$

Then $r_k = b - A x_k = \beta_1 u_1 - A V_k y_k = U_{k+1} e_1 \beta_1 - U_{k+1} B_k y_k$ gives

$$\begin{aligned} r_k &= U_{k+1} \left( e_1 \beta_1 - Q_{k+1}^T \begin{pmatrix} R_k \\ 0 \end{pmatrix} y_k \right) = U_{k+1} \left( e_1 \beta_1 - Q_{k+1}^T \begin{pmatrix} t_k \\ 0 \end{pmatrix} \right) \\ &= U_{k+1} \left( Q_{k+1}^T \begin{pmatrix} \widetilde{Q}_k^T & \\ & 1 \end{pmatrix} \tilde{b}_k - Q_{k+1}^T \begin{pmatrix} \widetilde{Q}_k^T \tilde{t}_k \\ 0 \end{pmatrix} \right) \\ &= U_{k+1} Q_{k+1}^T \begin{pmatrix} \widetilde{Q}_k^T & \\ & 1 \end{pmatrix} \left( \tilde{b}_k - \begin{pmatrix} \tilde{t}_k \\ 0 \end{pmatrix} \right). \end{aligned}$$

Therefore, assuming orthogonality of $U_{k+1}$, we have

$$\|r_k\| = \left\| \tilde{b}_k - \begin{pmatrix} \tilde{t}_k \\ 0 \end{pmatrix} \right\|. \tag{3.2}$$

The vectors $\tilde{b}_k$ and $\tilde{t}_k$ can be written in the form

$$\tilde{b}_k = \begin{pmatrix} \tilde{\beta}_1 & \cdots & \tilde{\beta}_{k-1} & \dot{\beta}_k & \ddot{\beta}_{k+1} \end{pmatrix}^T \qquad \tilde{t}_k = \begin{pmatrix} \tilde{\tau}_1 & \cdots & \tilde{\tau}_{k-1} & \dot{\tau}_k \end{pmatrix}^T. \tag{3.3}$$

The vector $\tilde{t}_k$ can be computed by forward substitution from $\widetilde{R}_k^T \tilde{t}_k = z_k$.

Lemma 3.1. *In (3.2)–(3.3), $\tilde{\beta}_i = \tilde{\tau}_i$ for $i = 1, \dots, k-1$.*

*Proof.* Appendix A proves the lemma by induction. □

Using this lemma we can estimate $\|r_k\|$ from just the last two elements of $\tilde{b}_k$ and the last element of $\tilde{t}_k$, as shown in (3.6) below.

**3.1.1. Pseudo-code for computing $\|r_k\|$.** The following summarizes how $\|r_k\|$ may be obtained from quantities arising from the first and third QR factorizations.

1. (Initialize)

$$\ddot{\beta}_1 = \beta_1 \qquad \dot{\beta}_0 = 0 \qquad \dot{\rho}_0 = 1 \qquad \tilde{\tau}_{-1} = 0 \qquad \tilde{\theta}_0 = 0 \qquad \zeta_0 = 0$$

2. For the $k$th iteration, repeat steps 3–6.
3. (Apply rotation $P_k$)

$$\hat{\beta}_k = c_k \ddot{\beta}_k \qquad \ddot{\beta}_{k+1} = -s_k \ddot{\beta}_k \tag{3.4}$$

4. (If $k \ge 2$, construct and apply rotation $\widetilde{P}_{k-1}$)

$$\begin{aligned} \tilde{\rho}_{k-1} &= \left(\dot{\rho}_{k-1}^2 + \bar{\theta}_k^2\right)^{\frac{1}{2}} & & \\ \tilde{c}_{k-1} &= \dot{\rho}_{k-1}/\tilde{\rho}_{k-1} & \tilde{s}_{k-1} &= \bar{\theta}_k/\tilde{\rho}_{k-1} \\ \tilde{\theta}_k &= \tilde{s}_{k-1}\bar{\rho}_k & \dot{\rho}_k &= \tilde{c}_{k-1}\bar{\rho}_k \\ \tilde{\beta}_{k-1} &= \tilde{c}_{k-1}\dot{\beta}_{k-1} + \tilde{s}_{k-1}\hat{\beta}_k & \dot{\beta}_k &= -\tilde{s}_{k-1}\dot{\beta}_{k-1} + \tilde{c}_{k-1}\hat{\beta}_k \end{aligned} \tag{3.5}$$

5. (Update $\tilde{t}_k$ by forward substitution)

$$\tilde{\tau}_{k-1} = (\zeta_{k-1} - \tilde{\theta}_{k-1}\tilde{\tau}_{k-2})/\tilde{\rho}_{k-1} \qquad \dot{\tau}_k = (\zeta_k - \tilde{\theta}_k \tilde{\tau}_{k-1})/\dot{\rho}_k$$

6. (Form $\|r_k\|$)

$$\gamma = (\dot{\beta}_k - \dot{\tau}_k)^2 + \ddot{\beta}_{k+1}^2, \qquad \|r_k\| = \sqrt{\gamma} \tag{3.6}$$

**3.2. Computing $\|A^T r_k\|$.** From (2.5) we have $\|A^T r_k\| = |\bar{\zeta}_{k+1}|$, which by (2.10) is monotonically decreasing.

**3.3. Computing $\|x_k\|$.** From section 2.4 we have $x_k = V_k R_k^{-1} \bar{R}_k^{-1} z_k$. From the third QR factorization $\widetilde{Q}_k \bar{R}_k^T = \widetilde{R}_k$ in section 3.1 and a fourth QR factorization $\hat{Q}_k (\widetilde{Q}_k R_k)^T = \hat{R}_k$ we can write

$$x_k = V_k R_k^{-1} \bar{R}_k^{-1} z_k = V_k R_k^{-1} \bar{R}_k^{-1} \bar{R}_k \widetilde{Q}_k^T \tilde{z}_k = V_k R_k^{-1} \widetilde{Q}_k^T \widetilde{Q}_k R_k \hat{Q}_k^T \hat{z}_k = V_k \hat{Q}_k^T \hat{z}_k,$$

where $\tilde{z}_k$ and $\hat{z}_k$ are defined by forward substitutions $\widetilde{R}_k^T \tilde{z}_k = z_k$ and $\hat{R}_k^T \hat{z}_k = \tilde{z}_k$. Assuming orthogonality of $V_k$ we arrive at the estimate $\|x_k\| = \|\hat{z}_k\|$. Since only the last diagonal of $\widetilde{R}_k$ and the bottom $2 \times 2$ part of $\hat{R}_k$ change each iteration, this estimate of $\|x_k\|$ can again be updated cheaply. The pseudo-code, omitted here, can be derived as in section 3.1.1. Experimentally we have observed that for every iteration, $\|x_k\| > \|x_{k-1}\|$ is either true or very nearly true.

**3.4. Estimates of $\|A\|$ and cond$(A)$.** It is known that the singular values of $B_k$ are interlaced by those of $A$ and are bounded above and below by the largest and smallest nonzero singular values of $A$ [16]. Therefore we can estimate $\|A\|$ and cond$(A)$ by $\|B_k\|$ and cond$(B_k)$ respectively. Considering the Frobenius norm of $B_k$, we have the recurrence relation $\|B_{k+1}\|_F^2 = \|B_k\|_F^2 + \alpha_k^2 + \beta_{k+1}^2$. From (2.3)–(2.4) and (2.6), we can show that the following QLP factorization [23] holds:

$$Q_{k+1} B_k \bar{Q}_k^T = \begin{pmatrix} \bar{R}_{k-1}^T & \\ \bar{\theta}_k e_{k-1}^T & \bar{c}_{k-1}\rho_k \end{pmatrix}$$

(the same as $\bar{R}_k^T$ except for the last diagonal). Since the singular values of $B_k$ are approximated by the diagonal elements of that lower-bidiagonal matrix [23], and since the diagonals are all positive, we can estimate cond$(A)$ by the ratio of the largest and smallest values in $\{\bar{\rho}_1, \ldots, \bar{\rho}_{k-1}, \bar{c}_{k-1}\rho_k\}$. Those values can be updated cheaply.

**3.5. Stopping criteria.** With exact arithmetic, the Golub-Kahan process terminates when either $\alpha_{k+1} = 0$ or $\beta_{k+1} = 0$. For certain data $b$, this could happen in practice when $k$ is small (but is unlikely later). We show that LSMR will have solved the problem at that point and should therefore terminate.

When $\alpha_{k+1} = 0$, with the expression of $\|A^T r_k\|$ from section 3.2, we have

$$\|A^T r_k\| = |\bar{\zeta}_{k+1}| = |\bar{s}_k \bar{\zeta}_k| = |\theta_{k+1} \bar{\rho}_k^{-1} \bar{\zeta}_k| = |s_k \alpha_{k+1} \bar{\rho}_k^{-1} \bar{\zeta}_k| = 0,$$

where (2.10), (2.9), (2.8) are used. Thus, a least-squares solution has been obtained.

When $\beta_{k+1} = 0$, we have

$$\begin{aligned}
s_k &= \beta_{k+1} \rho_k^{-1} = 0. && \text{(from (2.7))} \quad (3.7)\\
\ddot{\beta}_{k+1} &= -s_k \ddot{\beta}_k = 0. && \text{(from (3.4), (3.7))} \quad (3.8)\\
\dot{\beta}_k &= \tilde{c}_k^{-1} \left( \tilde{\beta}_k - \tilde{s}_k (-1)^k s^{(k)} c_{k+1} \beta_1 \right) && \text{(from (A.6))}\\
&= \tilde{c}_k^{-1} \tilde{\beta}_k && \text{(from (3.7))}\\
&= \dot{\rho}_k^{-1} \tilde{\rho}_k \tilde{\beta}_k && \text{(from (3.5))}\\
&= \dot{\rho}_k^{-1} \tilde{\rho}_k \tilde{\tau}_k && \text{(from Lemma 3.1)}\\
&= \dot{\tau}_k. && \text{(from (A.2), (A.3))} \quad (3.9)
\end{aligned}$$

By (3.9), (3.8), and (3.6) we conclude that $\|r_k\| = 0$. It follows that $Ax_k = b$.

**3.6. Practical stopping criteria.** For LSMR we use the same stopping rules as LSQR [16], involving dimensionless quantities ATOL, BTOL, CONLIM:

S1: Stop if $\|r_k\| \le \text{BTOL}\|b\| + \text{ATOL}\|A\|\|x_k\|$
S2: Stop if $\|A^T r_k\| \le \text{ATOL}\|A\|\|r_k\|$
S3: Stop if $\text{cond}(A) \ge \text{CONLIM}$

S1 applies to consistent systems, allowing for uncertainty in $A$ and $b$ [10, Theorem 7.1]. S2 applies to inconsistent systems and comes from Stewart's backward error estimate $\|E_2\|$ assuming uncertainty in $A$; see section 6.1. S3 applies to any system.

**4. Characteristics of the solution on singular systems.** If $A$ does not have full column rank, the normal equation $A^T Ax = A^T b$ is singular but consistent. We show that LSQR and LSMR both give the minimum-norm LS solution. That is, they both solve the optimization problem $\min \|x\|_2$ such that $A^T Ax = A^T b$. Let $\text{N}(A)$ and $\text{R}(A)$ denote the nullspace and range of a matrix $A$.

LEMMA 4.1. *If $A \in \mathbb{R}^{m\times n}$ and $p \in \mathbb{R}^n$ satisfy $A^T Ap = 0$, then $p \in N(A)$.*

*Proof.* $A^T Ap = 0 \Rightarrow p^T A^T Ap = 0 \Rightarrow (Ap)^T Ap = 0 \Rightarrow Ap = 0$. ∎

THEOREM 4.2. *LSQR returns the minimum-norm solution.*

*Proof.* The final LSQR solution satisfies $A^T Ax_k^{\text{LSQR}} = A^T b$, and any other solution $\widehat{x}$ satisfies $A^T A\widehat{x} = A^T b$. With $p = \widehat{x} - x_k^{\text{LSQR}}$, the difference between the two normal equations gives $A^T Ap = 0$, so that $Ap = 0$ by Lemma 4.1. From $\alpha_1 v_1 = A^T u_1$ and $\alpha_{k+1} v_{k+1} = A^T u_{k+1} - \beta_{k+1} v_k$ (2.1), we have $v_1, \dots, v_k \in \text{R}(A^T)$. With $Ap = 0$, this implies $p^T V_k = 0$, so that

$$\begin{aligned}
\|\widehat{x}\|_2^2 - \|x_k^{\text{LSQR}}\|_2^2 = \|x_k^{\text{LSQR}} + p\|_2^2 - \|x_k^{\text{LSQR}}\|_2^2 &= p^T p + 2p^T x_k^{\text{LSQR}}\\
&= p^T p + 2p^T V_k y_k^{\text{LSQR}} = p^T p \ge 0.
\end{aligned}$$

∎

COROLLARY 4.3. *LSMR returns the minimum-norm solution.*

*Proof.* At convergence, $\alpha_{k+1} = 0$ or $\beta_{k+1} = 0$. Thus $\bar{\beta}_{k+1} = \alpha_{k+1}\beta_{k+1} = 0$, which means equation (2.2) becomes $\min \|\bar{\beta}_1 e_1 - B_k^T B_k y_k\|$ and hence $B_k^T B_k y_k = \bar{\beta}_1 e_1$, since $B_k$ has full rank. This is the normal equation for $\min \|B_k y_k - \beta_1 e_1\|$, the same LS subproblem solved by LSQR. We conclude that at convergence, $y_k = y_k^{\mathrm{LSQR}}$ and thus $x_k = V_k y_k = V_k y_k^{\mathrm{LSQR}} = x_k^{\mathrm{LSQR}}$, and Theorem 4.2 applies. □

**4.1. Complexity.** We compare the storage requirement and computational complexity for LSMR and LSQR on $Ax \approx b$ and MINRES on the normal equation $A^T Ax = A^T b$. In Table 4.1, we list the vector storage needed (excluding storage for $A$ and $b$). Recall that $A$ is $m \times n$ and for LS systems $m$ may be considerably larger than $n$. $Av$ denotes the working storage for matrix-vector products. Work represents the number of floating-point multiplications required at each iteration.

TABLE 4.1
*Storage and computational requirements for various least-squares methods*

| | Storage | | Work | |
|---|---|---|---|---|
| | $m$ | $n$ | $m$ | $n$ |
| LSMR | $Av, u$ | $x, v, h, \bar{h}$ | 3 | 6 |
| LSQR | $Av, u$ | $x, v, w$ | 3 | 5 |
| MINRES on $A^T Ax = A^T b$ | $Av$ | $x, v_1, v_2, w_1, w_2, w_3$ | | 8 |

**5. Regularized least squares.** In this section, we extend LSMR to the regularized LS problem:

$$\min \left\| \begin{pmatrix} A \\ \lambda I \end{pmatrix} x - \begin{pmatrix} b \\ 0 \end{pmatrix} \right\|_2 . \tag{5.1}$$

If $\bar{A} = \begin{pmatrix} A \\ \lambda I \end{pmatrix}$ and $\bar{r}_k = \begin{pmatrix} b \\ 0 \end{pmatrix} - \bar{A}x_k$, then

$$\begin{aligned} \bar{A}^T \bar{r}_k = A^T r_k - \lambda^2 x_k &= V_{k+1} \left( \bar{\beta}_1 e_1 - \begin{pmatrix} B_k^T B_k \\ \bar{\beta}_{k+1} e_k^T \end{pmatrix} y_k - \lambda^2 \begin{pmatrix} y_k \\ 0 \end{pmatrix} \right) \\ &= V_{k+1} \left( \bar{\beta}_1 e_1 - \begin{pmatrix} R_k^T R_k \\ \bar{\beta}_{k+1} e_k^T \end{pmatrix} y_k \right) \end{aligned}$$

and the rest of the main algorithm follows the same as in the unregularized case. In the last equality, $R_k$ is defined by the QR factorization

$$Q_{2k+1} \begin{pmatrix} B_k \\ \lambda I \end{pmatrix} = \begin{pmatrix} R_k \\ 0 \end{pmatrix}, \qquad Q_{2k+1} \equiv P_k \hat{P}_k \dots P_2 \hat{P}_2 P_1 \hat{P}_1,$$

where $\hat{P}_l$ is a rotation operating on rows $l$ and $l + k + 1$. The effects of $\hat{P}_1$ and $P_1$ are illustrated here:

$$\hat{P}_1 \begin{pmatrix} \alpha_1 & \\ \beta_2 & \alpha_2 \\ & \beta_3 \\ \lambda & \\ & \lambda \end{pmatrix} = \begin{pmatrix} \hat{\alpha}_1 & \\ \beta_2 & \alpha_2 \\ & \beta_3 \\ 0 & \\ & \lambda \end{pmatrix}, \qquad P_1 \begin{pmatrix} \hat{\alpha}_1 & \\ \beta_2 & \alpha_2 \\ & \beta_3 \\ & \\ & \lambda \end{pmatrix} = \begin{pmatrix} \rho_1 & \theta_2 \\ & \bar{\alpha}_2 \\ & \beta_3 \\ & \\ & \lambda \end{pmatrix}.$$

**5.1. Effects on $\|\bar{r}_k\|$.** The introduction of regularization changes the residual norm as follows:

$$\begin{aligned}
\bar{r}_k = \begin{pmatrix} b \\ 0 \end{pmatrix} - \begin{pmatrix} A \\ \lambda I \end{pmatrix} x_k &= \begin{pmatrix} u_1 \\ 0 \end{pmatrix} \beta_1 - \begin{pmatrix} AV_k \\ \lambda V_k \end{pmatrix} y_k = \begin{pmatrix} u_1 \\ 0 \end{pmatrix} \beta_1 - \begin{pmatrix} U_{k+1}B_k \\ \lambda V_k \end{pmatrix} y_k \\
&= \begin{pmatrix} U_{k+1} & \\ & V_k \end{pmatrix} \left( e_1\beta_1 - \begin{pmatrix} B_k \\ \lambda I \end{pmatrix} y_k \right) \\
&= \begin{pmatrix} U_{k+1} & \\ & V_k \end{pmatrix} \left( e_1\beta_1 - Q_{2k+1}^T \begin{pmatrix} R_k \\ 0 \end{pmatrix} y_k \right) \\
&= \begin{pmatrix} U_{k+1} & \\ & V_k \end{pmatrix} \left( e_1\beta_1 - Q_{2k+1}^T \begin{pmatrix} t_k \\ 0 \end{pmatrix} \right) \\
&= \begin{pmatrix} U_{k+1} & \\ & V_k \end{pmatrix} Q_{2k+1}^T \begin{pmatrix} \widetilde{Q}_k^T & \\ & 1 \end{pmatrix} \left( \tilde{b}_k - \begin{pmatrix} \tilde{t}_k \\ 0 \end{pmatrix} \right)
\end{aligned}$$

with $\tilde{b}_k = \begin{pmatrix} \widetilde{Q}_k & \\ & 1 \end{pmatrix} Q_{2k+1} e_1 \beta_1$, where we adopt the notation

$$\tilde{b}_k = \begin{pmatrix} \tilde{\beta}_1 & \cdots & \tilde{\beta}_{k-1} & \dot{\beta}_k & \ddot{\beta}_{k+1} & \check{\beta}_1 & \cdots & \check{\beta}_k \end{pmatrix}^T.$$

We conclude that $\|\bar{r}_k\|^2 = \check{\beta}_1^2 + \cdots + \check{\beta}_k^2 + (\dot{\beta}_k - \tau_k)^2 + \ddot{\beta}_{k+1}^2$. The effect of regularization on the rotations is summarized as

$$\begin{pmatrix} \hat{c}_k & \hat{s}_k \\ -\hat{s}_k & \hat{c}_k \end{pmatrix} \begin{pmatrix} \bar{\alpha}_k & \ddot{\beta}_k \\ \lambda & \end{pmatrix} = \begin{pmatrix} \hat{\alpha}_k & \acute{\beta}_k \\ & \check{\beta}_k \end{pmatrix}$$

$$\begin{pmatrix} c_k & s_k \\ -s_k & c_k \end{pmatrix} \begin{pmatrix} \hat{\alpha}_k & & \acute{\beta}_k \\ \beta_{k+1} & \alpha_{k+1} & \end{pmatrix} = \begin{pmatrix} \rho_k & \theta_{k+1} & \hat{\beta}_k \\ & \bar{\alpha}_{k+1} & \ddot{\beta}_{k+1} \end{pmatrix}.$$

**5.2. Pseudo-code for regularized LSMR.** The following summarizes algorithm LSMR for solving the regularized problem (5.1) with given $\lambda$. Our MATLAB implementation is based on these steps.

1. (Initialize)

$$\begin{array}{cccccc}
\beta_1 u_1 = b & \alpha_1 v_1 = A^T u_1 & \bar{\alpha}_1 = \alpha_1 & \bar{\zeta}_1 = \alpha_1\beta_1 & \rho_0 = 1 & \bar{\rho}_0 = 1 \\
\bar{c}_0 = 1 & \bar{s}_0 = 0 & \ddot{\beta}_1 = \beta_1 & \dot{\beta}_0 = 0 & \dot{\rho}_0 = 1 & \tilde{\tau}_{-1} = 0 \\
\tilde{\theta}_0 = 0 & \zeta_0 = 0 & d_0 = 0 & h_1 = v_1 & \bar{h}_0 = 0 & x_0 = 0
\end{array}$$

2. For $k = 1, 2, 3, \ldots$ repeat steps 3–12.
3. (Continue the bidiagonalization)

$$\beta_{k+1} u_{k+1} = A v_k - \alpha_k u_k \qquad \alpha_{k+1} v_{k+1} = A^T u_{k+1} - \beta_{k+1} v_k$$

4. (Construct rotation $\hat{P}_k$)

$$\hat{\alpha}_k = \left(\bar{\alpha}_k^2 + \lambda^2\right)^{\frac{1}{2}} \qquad \hat{c}_k = \bar{\alpha}_k / \hat{\alpha}_k \qquad \hat{s}_k = \lambda / \hat{\alpha}_k$$

5. (Construct and apply rotation $P_k$)

$$\begin{array}{lll}
\rho_k = \left(\hat{\alpha}_k^2 + \beta_{k+1}^2\right)^{\frac{1}{2}} & c_k = \hat{\alpha}_k / \rho_k & s_k = \beta_{k+1} / \rho_k \\
\theta_{k+1} = s_k \alpha_{k+1} & \bar{\alpha}_{k+1} = c_k \alpha_{k+1} &
\end{array}$$

6. (Construct and apply rotation $\bar{P}_k$)

$$\begin{aligned} \bar{\theta}_k &= \bar{s}_{k-1}\rho_k & \bar{\rho}_k &= \left((\bar{c}_{k-1}\rho_k)^2 + \theta_{k+1}^2\right)^{\frac{1}{2}} \\ \bar{c}_k &= \bar{c}_{k-1}\rho_k/\bar{\rho}_k & \bar{s}_k &= \theta_{k+1}/\bar{\rho}_k \\ \zeta_k &= \bar{c}_k\bar{\zeta}_k & \bar{\zeta}_{k+1} &= -\bar{s}_k\bar{\zeta}_k \end{aligned}$$

7. (Update $\bar{h}$, $x$, $h$)

$$\begin{aligned} \bar{h}_k &= h_k - (\bar{\theta}_k\rho_k/(\rho_{k-1}\bar{\rho}_{k-1}))\bar{h}_{k-1} \\ x_k &= x_{k-1} + (\zeta_k/(\rho_k\bar{\rho}_k))\bar{h}_k \\ h_{k+1} &= v_{k+1} - (\theta_{k+1}/\rho_k)h_k \end{aligned}$$

8. (Apply rotation $\hat{P}_k, P_k$)

$$\acute{\beta}_k = \hat{c}_k\ddot{\beta}_k \qquad \check{\beta}_k = -\hat{s}_k\ddot{\beta}_k \qquad \hat{\beta}_k = c_k\acute{\beta}_k \qquad \ddot{\beta}_{k+1} = -s_k\acute{\beta}_k$$

9. (If $k \geq 2$, construct and apply rotation $\widetilde{P}_{k-1}$)

$$\begin{aligned} \tilde{\rho}_{k-1} &= \left(\dot{\rho}_{k-1}^2 + \bar{\theta}_k^2\right)^{\frac{1}{2}} \\ \tilde{c}_{k-1} &= \dot{\rho}_{k-1}/\tilde{\rho}_{k-1} & \tilde{s}_{k-1} &= \bar{\theta}_k/\tilde{\rho}_{k-1} \\ \tilde{\theta}_k &= \tilde{s}_{k-1}\bar{\rho}_k & \dot{\rho}_k &= \tilde{c}_{k-1}\bar{\rho}_k \\ \tilde{\beta}_{k-1} &= \tilde{c}_{k-1}\dot{\beta}_{k-1} + \tilde{s}_{k-1}\hat{\beta}_k & \dot{\beta}_k &= -\tilde{s}_{k-1}\dot{\beta}_{k-1} + \tilde{c}_{k-1}\hat{\beta}_k \end{aligned}$$

10. (Update $\tilde{t}_k$ by forward substitution)

$$\tilde{\tau}_{k-1} = (\zeta_{k-1} - \tilde{\theta}_{k-1}\tilde{\tau}_{k-2})/\tilde{\rho}_{k-1} \qquad \dot{\tau}_k = (\zeta_k - \tilde{\theta}_k\tilde{\tau}_{k-1})/\dot{\rho}_k$$

11. (Compute $\|\bar{r}_k\|$)

$$d_k = d_{k-1} + \check{\beta}_k^2 \qquad \gamma = d_k + (\dot{\beta}_k - \dot{\tau}_k)^2 + \ddot{\beta}_{k+1}^2 \qquad \|\bar{r}_k\| = \sqrt{\gamma}$$

12. (Compute $\|\bar{A}^T\bar{r}_k\|$, $\|x_k\|$, estimate $\|\bar{A}\|$, cond($\bar{A}$), and test for termination)

   $\|\bar{A}^T\bar{r}_k\| = |\bar{\zeta}_{k+1}|$ (section 3.2)
   Compute $\|x_k\|$ (section 3.3)
   Estimate $\sigma_{\max}(B_k), \sigma_{\min}(B_k)$ and hence $\|\bar{A}\|, \mathrm{cond}(\bar{A})$ (section 3.4)
   Terminate if any of the stopping criteria are satisfied (section 3.6)

**6. Backward errors.** For inconsistent problems with uncertainty in $A$ (but not $b$), let $x$ be any approximate solution. The *normwise backward error* for $x$ measures the perturbation to $A$ that would make $x$ an exact LS solution:

$$\mu(x) \equiv \min_E \|E\| \quad \text{s.t.} \quad (A+E)^T(A+E)x = (A+E)^T b. \tag{6.1}$$

It is known to be the smallest singular value of a certain $m \times (n+m)$ matrix $C$; see Waldén et al. [26] and Higham [10, pp. 392–393]:

$$\mu(x) = \sigma_{\min}(C), \qquad C \equiv \left[A \quad \frac{\|r\|}{\|x\|}\left(I - \frac{rr^T}{\|r\|^2}\right)\right].$$

Since it is generally too expensive to evaluate $\mu(x)$, we need to find approximations.

**6.1. Approximate backward errors $E_1$ and $E_2$.** In 1975, Stewart [21] discussed a particular backward error estimate that we will call $E_1$. Let $\widehat{x}$ and $\widehat{r} = b - A\widehat{x}$ be the exact LS solution and residual. Stewart showed that an approximate solution $x$ with residual $r = b - Ax$ is the exact LS solution of the perturbed problem $\min \|b - (A + E_1)x\|$, where $E_1$ is the rank-one matrix

$$E_1 = \frac{ex^T}{\|x\|^2}, \qquad \|E_1\| = \frac{\|e\|}{\|x\|}, \qquad e \equiv r - \widehat{r}, \tag{6.2}$$

with $\|r\|^2 = \|\widehat{r}\|^2 + \|e\|^2$. Soon after, Stewart [22] gave a further important result that can be used within any LS solver. The approximate $x$ and a certain vector $\tilde{r} = b - (A + E_2)x$ are the exact solution and residual of the perturbed LS problem $\min \|b - (A + E_2)x\|$, where

$$E_2 = -\frac{rr^TA}{\|r\|^2}, \qquad \|E_2\| = \frac{\|A^Tr\|}{\|r\|}, \qquad r = b - Ax. \tag{6.3}$$

LSQR and LSMR both compute $\|E_2\|$ for each iterate $x_k$ because the current $\|r_k\|$ and $\|A^Tr_k\|$ can be accurately estimated at almost no cost. An added feature is that for both solvers, $\tilde{r} = b - (A + E_2)x_k = r_k$ because $E_2x_k = 0$ (assuming orthogonality of $V_k$). That is, $x_k$ and $r_k$ are theoretically exact for the perturbed LS problem $(A + E_2)x \approx b$.

Stopping rule S2 (section 3.6) requires $\|E_2\| \le \text{ATOL}\|A\|$. Hence the following property gives LSMR an advantage over LSQR for stopping early.

THEOREM 6.1. $\|E_2^{LSMR}\| \le \|E_2^{LSQR}\|$.

*Proof.* This follows from $\|A^Tr_k^{\text{LSMR}}\| \le \|A^Tr_k^{\text{LSQR}}\|$ and $\|r_k^{\text{LSMR}}\| \ge \|r_k^{\text{LSQR}}\|$. ☐

**6.2. Approximate optimal backward error $\widetilde{\mu}(x)$.** Various authors have derived expressions for a quantity $\widetilde{\mu}(x)$ that has proved to be a very accurate approximation to $\mu(x)$ in (6.1) when $x$ is at least moderately close to the exact solution $\widehat{x}$. Grcar, Saunders, and Su [24, 8] show that $\widetilde{\mu}(x)$ can be obtained from a full-rank LS problem as follows:

$$K = \begin{bmatrix} A \\ \frac{\|r\|}{\|x\|}I \end{bmatrix}, \qquad v = \begin{bmatrix} r \\ 0 \end{bmatrix}, \qquad \min_y \|Ky - v\|, \qquad \widetilde{\mu}(x) = \|Ky\|/\|x\|, \tag{6.4}$$

and give the following MATLAB script for computing the "economy size" sparse QR factorization $K = QR$ and $c \equiv Q^Tv$ (for which $\|c\| = \|Ky\|$) and thence $\widetilde{\mu}(x)$:

```
[m,n] = size(A);        r   = b - A*x;
normx = norm(x);        eta = norm(r)/normx;
p = colamd(A);
K = [A(:,p);  eta*speye(n)];
v = [    r ;    zeros(n,1)];
[c,R] = qr(K,v,0);      mutilde = norm(c)/normx;
```

In our experiments we use this script to compute $\widetilde{\mu}(x_k)$ for each LSQR and LSMR iterate $x_k$. We refer to this as the *optimal* backward error for $x_k$ because it is provably very close to the true $\mu(x_k)$ [7].

**6.3. Related work.** More precise stopping rules have been derived recently by Arioli and Gratton [1] and Titley-Péloquin et al. [3, 13, 25]. The rules allow for uncertainty in both $A$ and $b$, and may prove to be useful for LSQR, LSMR, and least-squares methods in general. However, we would like to emphasize that rule S2 already terminates LSMR significantly sooner than LSQR on most of our inconsistent test cases; see Theorem 6.1, Fig. 7.2 (left), and Fig. 7.3 (top left).

**7. Numerical results.** For test examples, we have drawn from the University of Florida Sparse Matrix Collection (Davis [5]). We discuss overdetermined systems first, and then some square examples.

**7.1. Least-squares problems.** The `LPnetlib` group provides data for 138 linear programming problems of widely varying origin, structure, and size. The constraint matrix and objective function may be used to define a sparse LS problem $\min \|Ax-b\|$. Each example was downloaded in MATLAB format, and a sparse matrix $A$ and dense vector $b$ were extracted from the data structure via `A = (Problem.A)'` and `b = Problem.c` (where `'` denotes transpose).

Five examples had $b = 0$, and a further six gave $A^Tb = 0$. The remaining 127 problems had up to 243000 rows, 10000 columns, and 1.4M nonzeros in $A$. Diagonal scaling was applied to the columns of $\begin{bmatrix} A & b \end{bmatrix}$ to give a scaled problem $\min \|Ax - b\|$ in which the columns of $A$ (and also $b$) have unit 2-norm. LSQR and LSMR were run on each of the 127 scaled problems with stopping tolerance ATOL $= 10^{-8}$, generating sequences of approximate solutions $\{x_k^{\text{LSQR}}\}$ and $\{x_k^{\text{LSMR}}\}$. The iteration indices $k$ are omitted below. The associated residual vectors are denoted by $r$ without ambiguity, and $x^*$ is the solution to the LS problem, or the minimum-norm solution to the LS problem if the system is singular.

As expected, the optimal residual is nonzero in all cases. We record some general observations.

1. $\|r^{\text{LSQR}}\|$ is monotonic by design. $\|r^{\text{LSMR}}\|$ seems to be monotonic (no counterexamples were found) and *nearly* as small as $\|r^{\text{LSQR}}\|$ for all iterations on almost all problems. Figure 7.1 shows a typical example and a rare case.

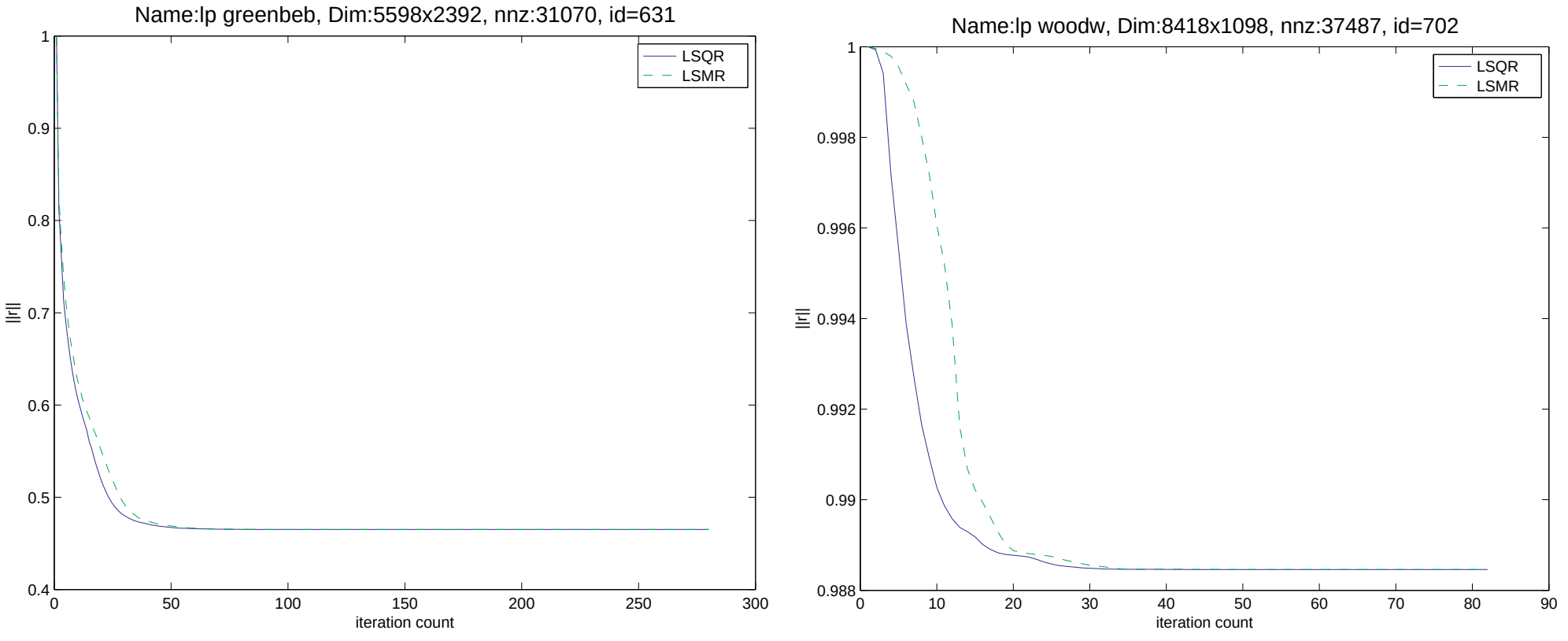

FIG. 7.1. *For most iterations, $\|r^{LSMR}\|$ appears to be monotonic and nearly as small as $\|r^{LSQR}\|$. Left: A typical case (problem* `lp_greenbeb`*). Right: A rare case (problem* `lp_woodw`*). LSMR's residual norm is significantly larger than LSQR's during early iterations.*

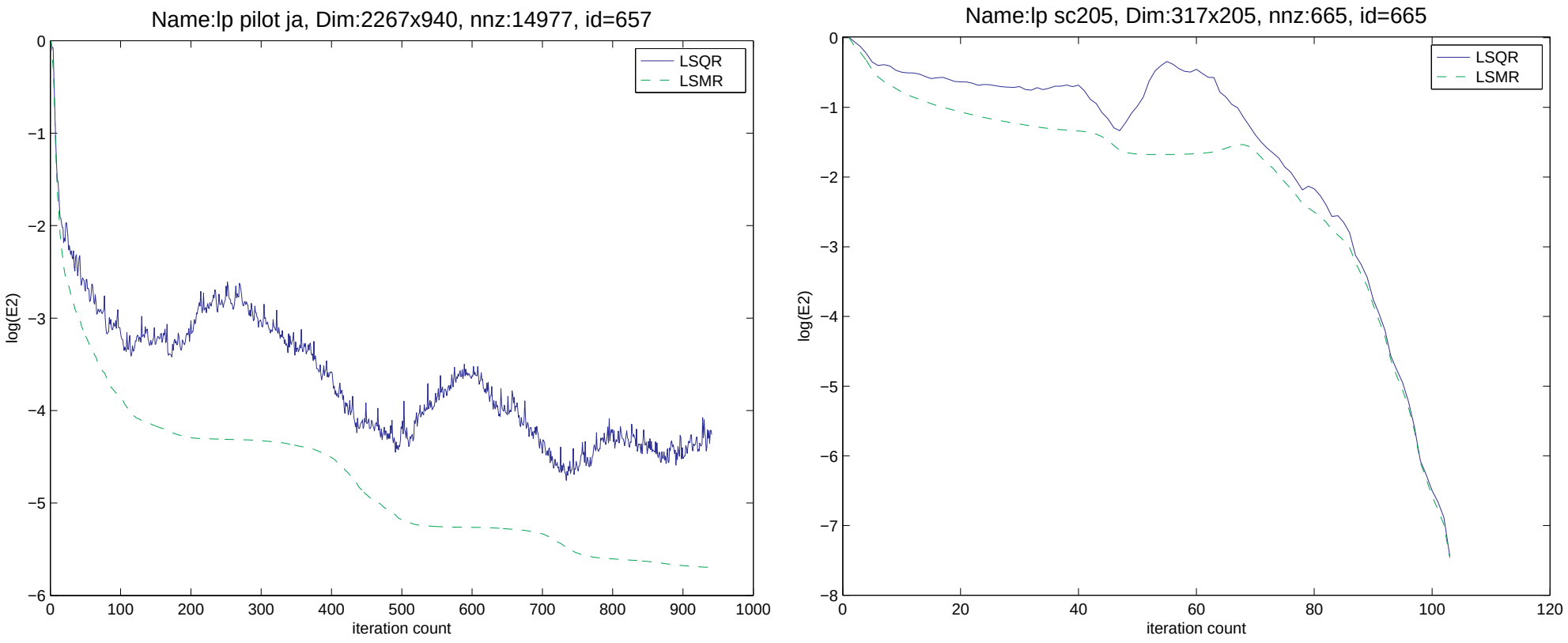

Fig. 7.2. *For most iterations, $\|E_2^{LSMR}\|$ appears to be monotonic (but $\|E_2^{LSQR}\|$ is not). Left: A typical case (problem* `lp_pilot_ja`*). LSMR is likely to terminate much sooner than LSMR (see Theorem 6.1). Right: Sole exception (problem* `lp_sc205`*) at iterations 54–67. The exception remains even if $U_k$ and/or $V_k$ are reorthogonalized.*

2. $\|x\|$ is nearly monotonic for LSQR and even more closely monotonic for LSMR. With $\|r\|$ monotonic for LSQR and essentially so for LSMR, $\|E_1\|$ in (6.2) is likely to appear monotonic for both solvers. Although $\|E_1\|$ is not normally available for each iteration, it provides a benchmark for $\|E_2\|$.
3. $\|E_2^{\text{LSQR}}\|$ is *not* monotonic, but $\|E_2^{\text{LSMR}}\|$ appears monotonic almost always. Figure 7.2 shows a typical case. The sole exception for this observation is also shown.
4. Note that Benbow [2] has given numerical results comparing a generalized form of LSQR with application of MINRES to the corresponding normal equation. The curves in [2, Figure 3] show the irregular and smooth behavior of LSQR and MINRES respectively in terms of $\|A^T r_k\|$. Those curves are effectively a preview of the left-hand plots in Figure 7.2 (where LSMR serves as our more reliable implementation of MINRES).
5. $\|E_1^{\text{LSQR}}\| \leq \|E_2^{\text{LSQR}}\|$ often, but not so for LSMR. Some examples are shown on Figure 7.3 along with $\widetilde{\mu}(x_k)$, the accurate estimate (6.4) of the optimal backward error for each point $x_k$.
6. $\|E_2^{\text{LSMR}}\| \approx \widetilde{\mu}(x^{\text{LSMR}})$ almost always. Figure 7.4 shows a typical example and a rare case. In all such "rare" cases, $\|E_1^{\text{LSMR}}\| \approx \widetilde{\mu}(x^{\text{LSMR}})$ instead!
7. $\widetilde{\mu}(x^{\text{LSQR}})$ is not always monotonic. $\widetilde{\mu}(x^{\text{LSMR}})$ does seem to be monotonic. Figure 7.5 gives examples.
8. $\widetilde{\mu}(x^{\text{LSMR}}) \leq \widetilde{\mu}(x^{\text{LSQR}})$ almost always. Figure 7.6 gives examples.
9. The errors $\|x^* - x^{\text{LSQR}}\|$ and $\|x^* - x^{\text{LSMR}}\|$ seem to decrease monotonically, with the LSQR error typically smaller than for LSMR. Figure 7.7 gives examples. This is one property for which LSQR seems more desirable (and it has been suggested [18] that for LS problems, LSQR could be terminated when rule S2 would terminate LSMR).

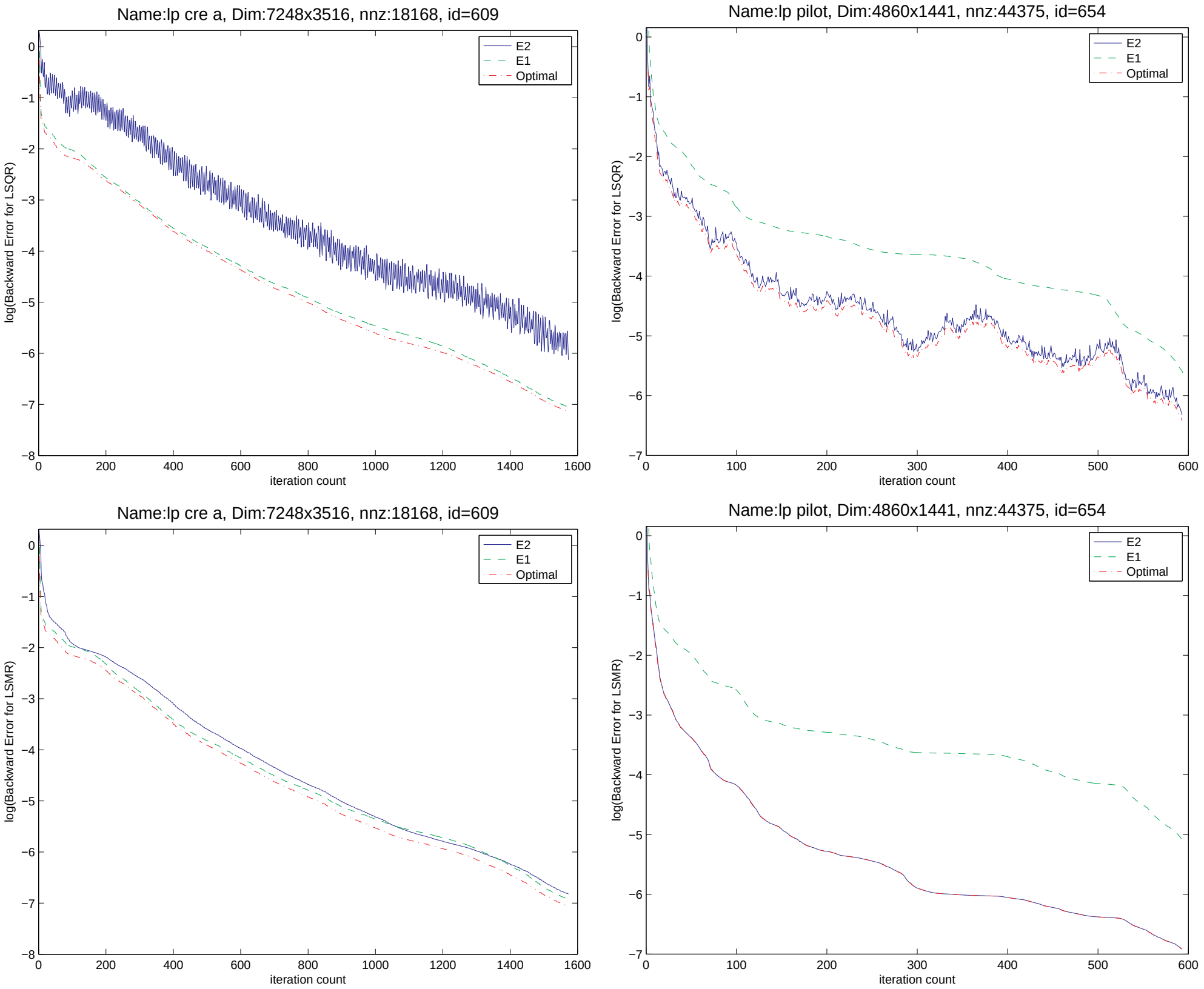

FIG. 7.3. *$\|E_1\|$, $\|E_2\|$, and $\widetilde{\mu}(x_k)$ for LSQR (top figures) and LSMR (bottom figures). Top left: A typical case. $\|E_1^{LSQR}\|$ is close to the optimal backward error, but the computable $\|E_2^{LSQR}\|$ is not. Top right: A rare case in which $\|E_2^{LSQR}\|$ is close to optimal. Bottom left: $\|E_1^{LSMR}\|$ and $\|E_2^{LSMR}\|$ are often both close to the optimal backward error. Bottom right: $\|E_1^{LSMR}\|$ is far from optimal, but the computable $\|E_2^{LSMR}\|$ is almost always close (too close to distinguish in the plot!). Problems* `lp_cre_a` *(left) and* `lp_pilot` *(right).*

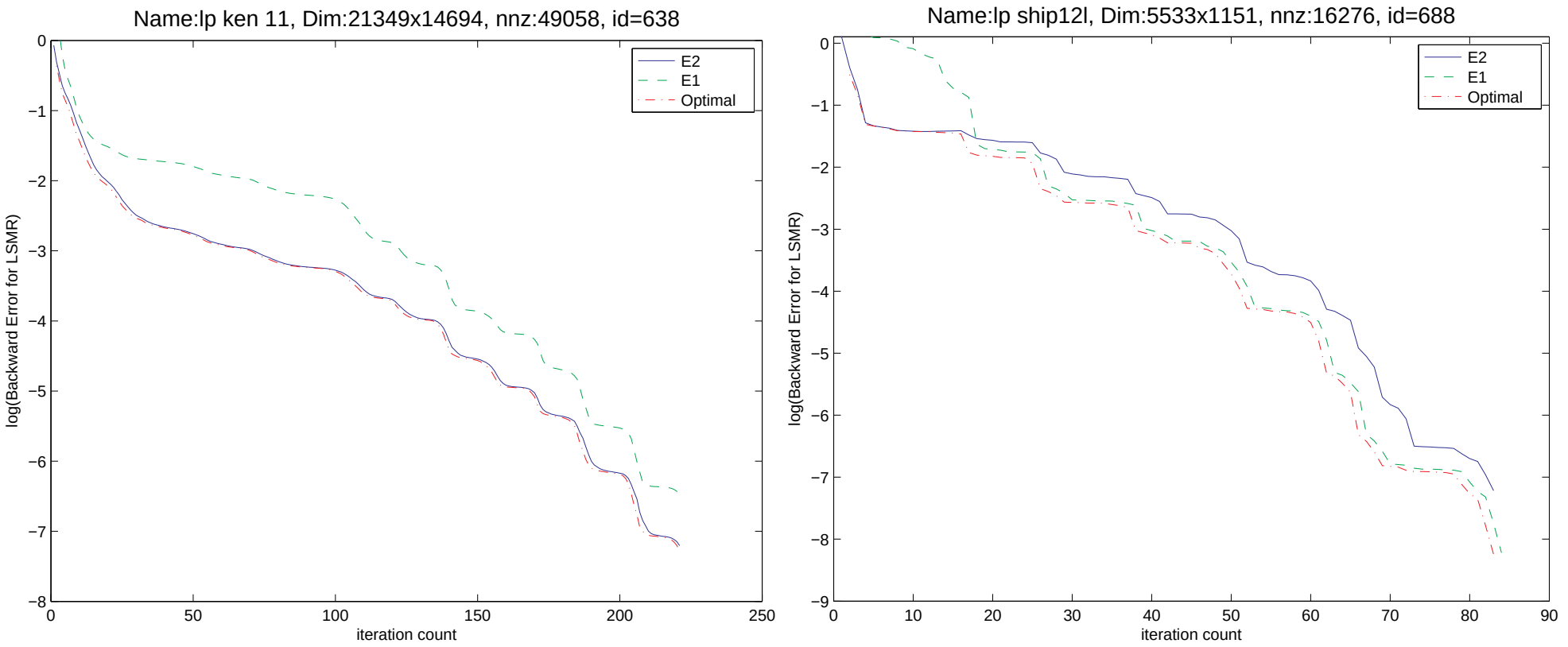

FIG. 7.4. *Again, $\|E_2^{LSMR}\| \approx \widetilde{\mu}(x^{LSMR})$ almost always (the computable backward error estimate is essentially optimal). Left: A typical case (problem* `lp_ken_11`*). Right: A rare case (problem* `lp_ship12l`*). Here, $\|E_1^{LSMR}\| \approx \widetilde{\mu}(x^{LSMR})$!*

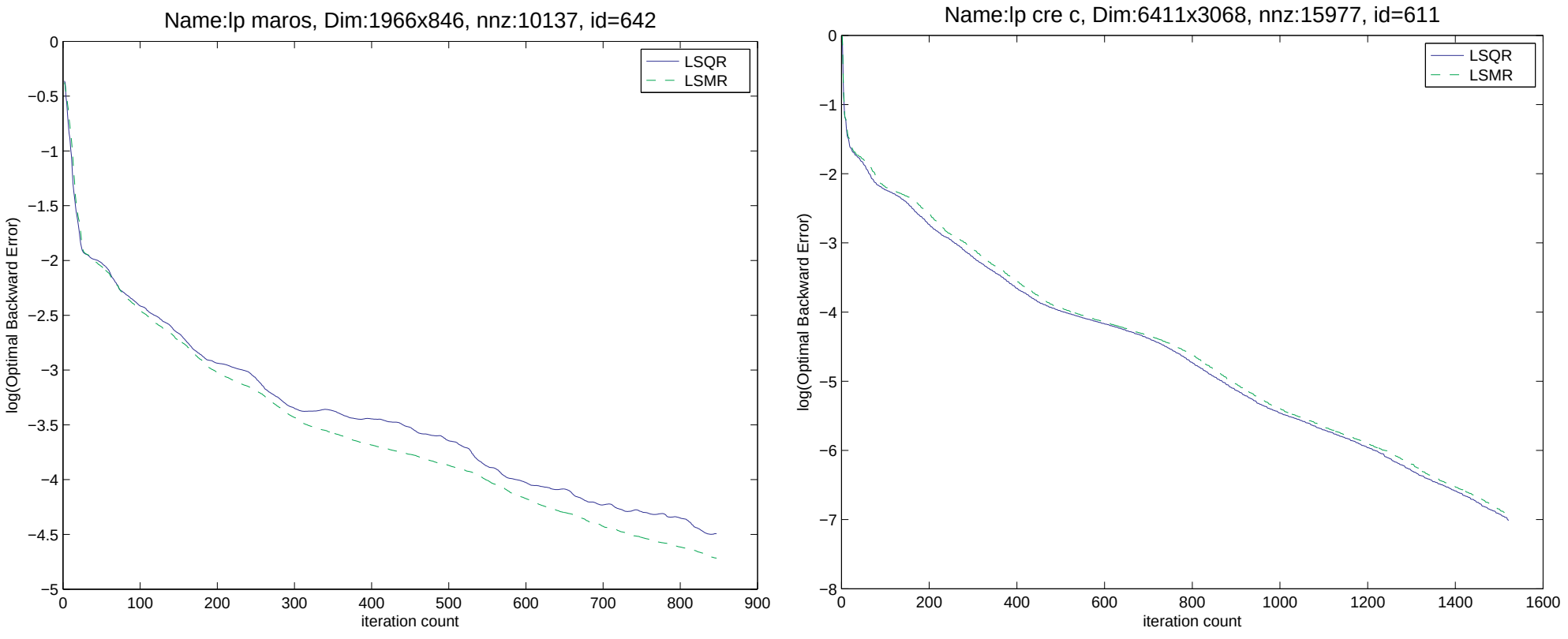

FIG. 7.5. *$\widetilde{\mu}(x^{LSMR})$ seems to be always monotonic, but $\widetilde{\mu}(x^{LSQR})$ is usually not. Left: A typical case for both LSQR and LSMR (problem* `lp_maros`*). Right: A rare case for LSQR, typical for LSMR (problem* `lp_cre_c`*).*

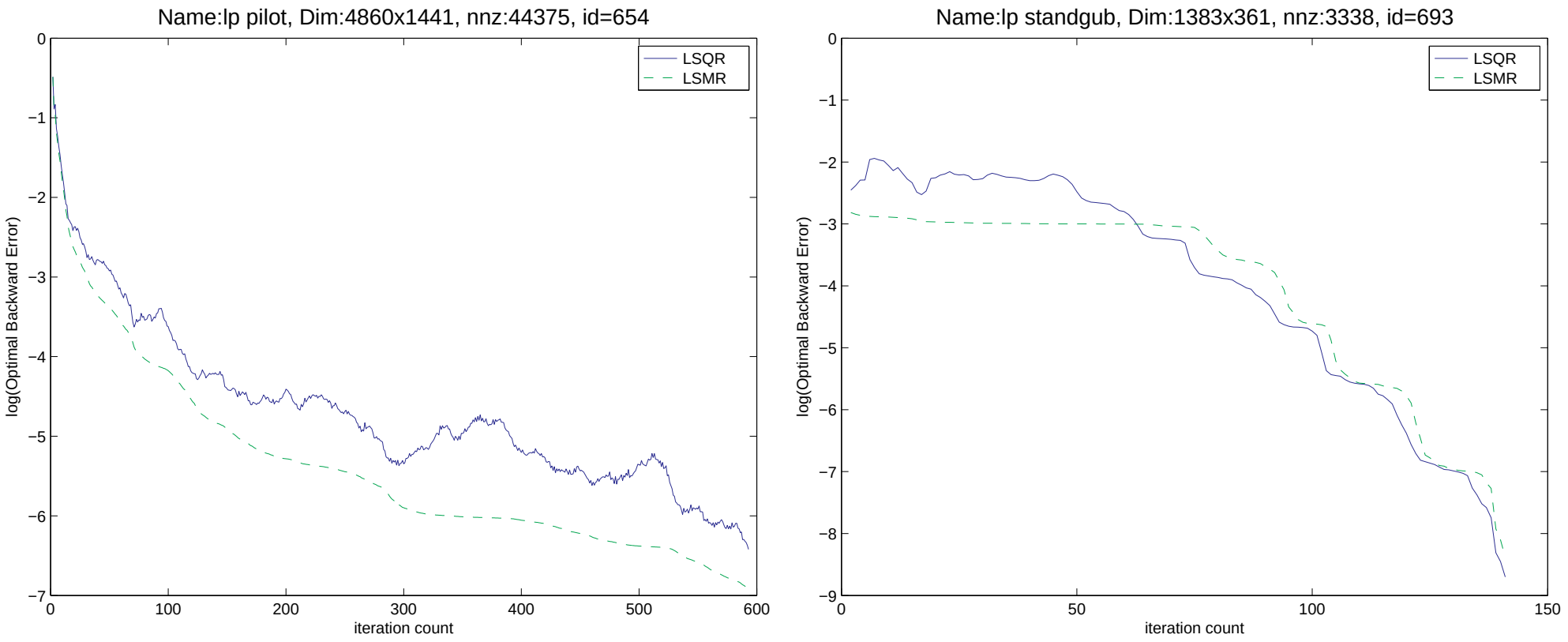

FIG. 7.6. *$\widetilde{\mu}(x^{LSMR}) \leq \widetilde{\mu}(x^{LSQR})$ almost always. Left: A typical case (problem* `lp_pilot`*). Right: A rare case (problem* `lp_standgub`*).*

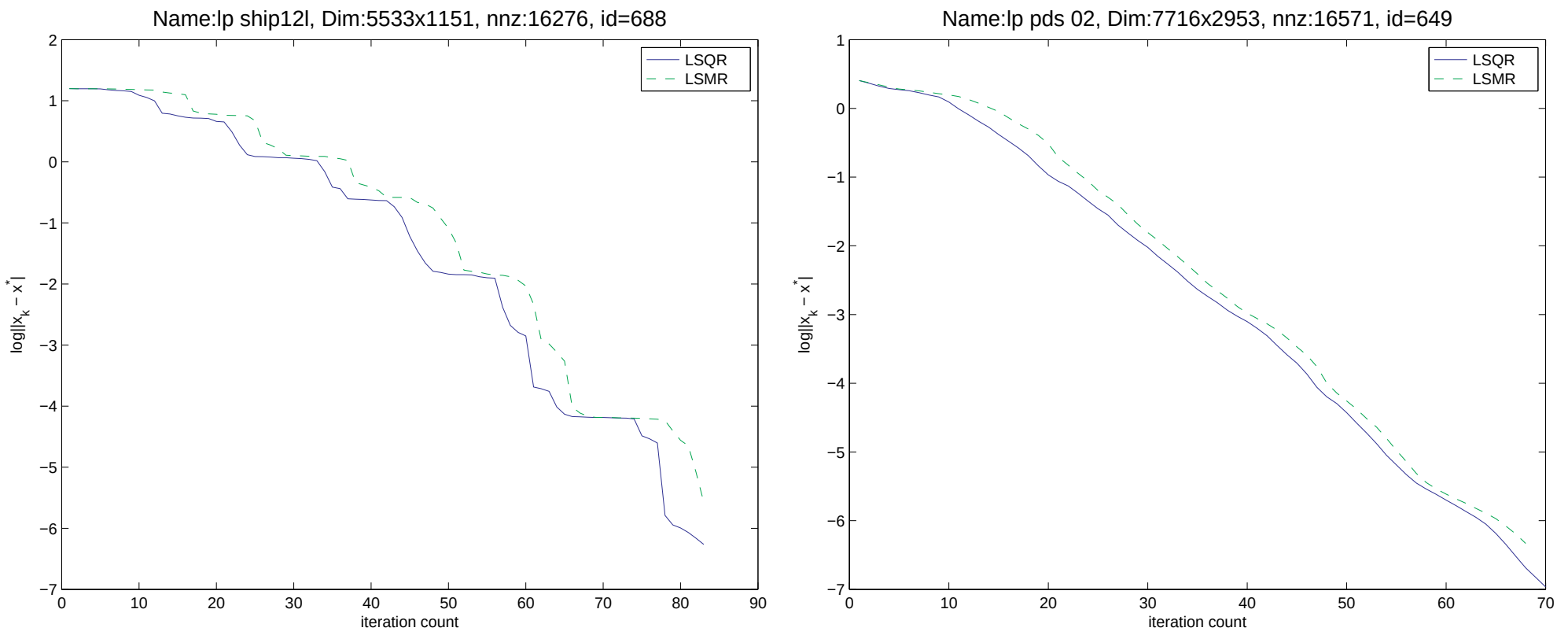

FIG. 7.7. *The errors $\|x^* - x^{LSQR}\|$ and $\|x^* - x^{LSMR}\|$ seem to decrease monotonically, with LSQR's errors smaller than for LSMR. Left: A nonsingular LS system (problem* `lp_ship12l`*). Right: A singular system (problem* `lp_pds_02`*). LSQR and LSMR both converge to the minimum-norm LS solution.*

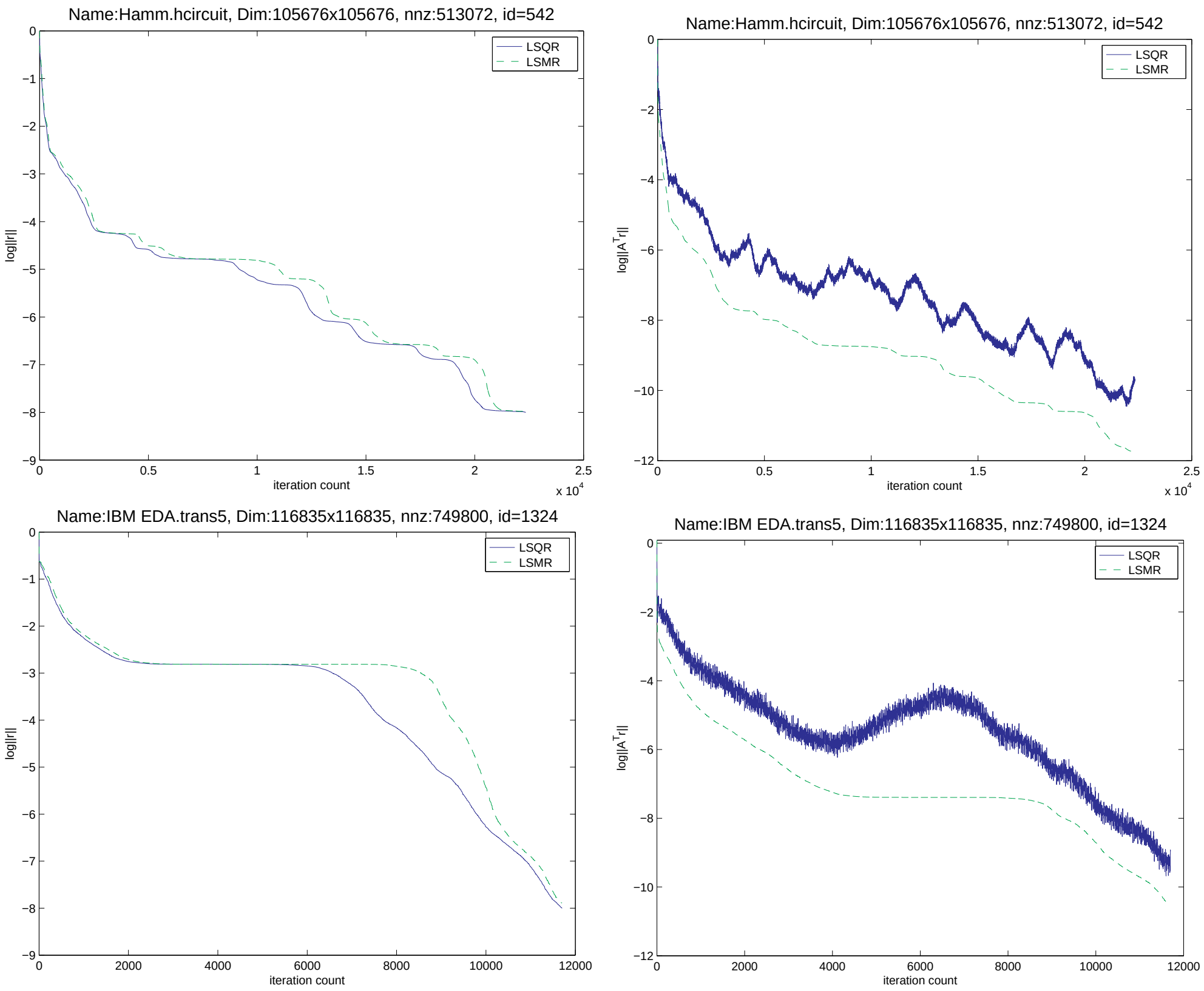

FIG. 7.8. *LSQR and LSMR solving two square nonsingular systems $Ax = b$: problems Hamm/hcircuit (top) and IBM_EDA/trans5 (bottom). Left:* $\log_{10} \|r_k\|$ *for both solvers, with prolonged plateaus for LSMR. Right:* $\log_{10} \|A^T r_k\|$ *(preferable for LSMR).*

**7.2. Square systems.** Since LSQR and LSMR are applicable to consistent systems, it is of interest to compare them on an unbiased test set. We used the search facility of Davis [5] to select a set of square real linear systems $Ax = b$. With `index = UFget`, the criteria

```
ids = find(index.nrows > 100000       & index.nrows < 200000 & ...
           index.nrows == index.ncols & index.isReal == 1    & ...
           index.posdef == 0          & index.numerical_symmetry < 1);
```

returned a list of 42 examples. Testing `isfield(UFget(id),'b')` left 26 cases for which $b$ was supplied. For each, diagonal scaling was first applied to the *rows* of $\begin{bmatrix} A & b \end{bmatrix}$ and then to its columns to give a scaled problem $Ax = b$ in which the columns of $\begin{bmatrix} A & b \end{bmatrix}$ have unit 2-norm. In spite of the scaling, most examples required more than $n$ iterations of LSQR or LSMR to reduce $\|r_k\|$ satisfactorily (rule S1 in section 3.6 with ATOL = BTOL = $10^{-8}$). To simulate better preconditioning, we chose two cases that required about $n/5$ and $n/10$ iterations. Figure 7.8 (left) shows both solvers reducing $\|r_k\|$ monotonically but with plateaus that are prolonged for LSMR. With loose stopping tolerances, LSQR could terminate somewhat sooner. Figure 7.8 (right) shows $\|A^T r_k\|$ for each solver. The plateaus for LSMR correspond to LSQR gaining ground with $\|r_k\|$, but falling significantly backward by the $\|A^T r_k\|$ measure.

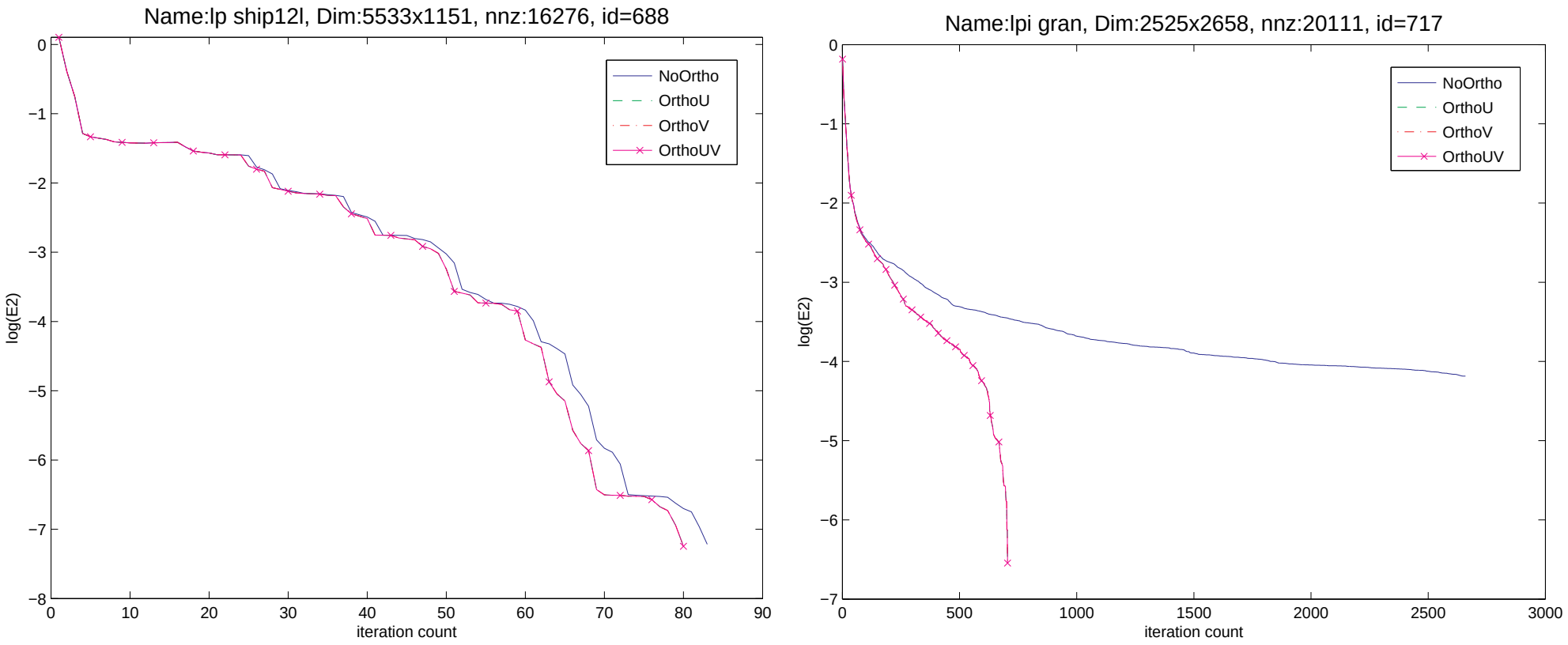

FIG. 7.9. *LSMR with and without reorthogonalization of $V_k$ and/or $U_k$. Left: An easy case where all options perform similarly (problem* `lp_ship12l`*). Right: A helpful case (problem* `lp_gran`*).*

**7.3. Reorthogonalization.** It is well known that Krylov-subspace methods can take arbitrarily many iterations because of loss of orthogonality. For the Golub-Kahan bidiagonalization, we have two sets of vectors $U_k$ and $V_k$. As an experiment, we implemented the following options in LSMR:

1. No reorthogonalization.
2. Reorthogonalize $V_k$ (that is, reorthogonalize $v_k$ with respect to $V_{k-1}$).
3. Reorthogonalize $U_k$ (that is, reorthogonalize $u_k$ with respect to $U_{k-1}$).
4. Both 2 and 3.

Each option was tested on all of the over-determined test problems with fewer than 16K nonzeros. Figure 7.9 shows an "easy" case in which all options converge equally well (convergence before significant loss of orthogonality), and an extreme case in which reorthogonalization makes a large difference.

Unexpectedly, options 2, 3, and 4 proved to be indistinguishable in all cases. To look closer, we forced LSMR to take $n$ iterations. Option 2 (with $V_k$ orthonormal to machine precision $\epsilon$) was found to be keeping $U_k$ orthonormal to at least $O(\sqrt{\epsilon})$. Option 3 (with $U_k$ orthonormal) was not quite as effective but it kept $V_k$ orthonormal to at least $O(\sqrt{\epsilon})$ up to the point where LSMR would terminate when ATOL $= \sqrt{\epsilon}$.

Note that for square or rectangular $A$ with exact arithmetic, LSMR is equivalent to MINRES on the normal equation (and hence to the conjugate-residual method [12] and GMRES [20] on the same equation). Reorthogonalization makes the equivalence essentially true in practice. We now focus on reorthogonalizing $V_k$ but not $U_k$.

Other authors have presented numerical results involving reorthogonalization. For example, on some randomly generated LS problems of increasing condition number, Hayami *et al.* [9] compare their BA-GMRES method with an implementation of CGLS (equivalent to LSQR [16]) in which $V_k$ is reorthogonalized, and find that the methods require essentially the same number of iterations. The preconditioner chosen for BA-GMRES made that method equivalent to GMRES on $A^TAx = A^Tb$. Thus, GMRES without reorthogonalization was seen to converge essentially as well as CGLS or LSQR with reorthogonalization of $V_k$ (option 2 above). This coincides with the analysis by Paige *et al.* [14], who conclude that MGS-GMRES does not need reorthogonalization of the Arnoldi vectors $V_k$.

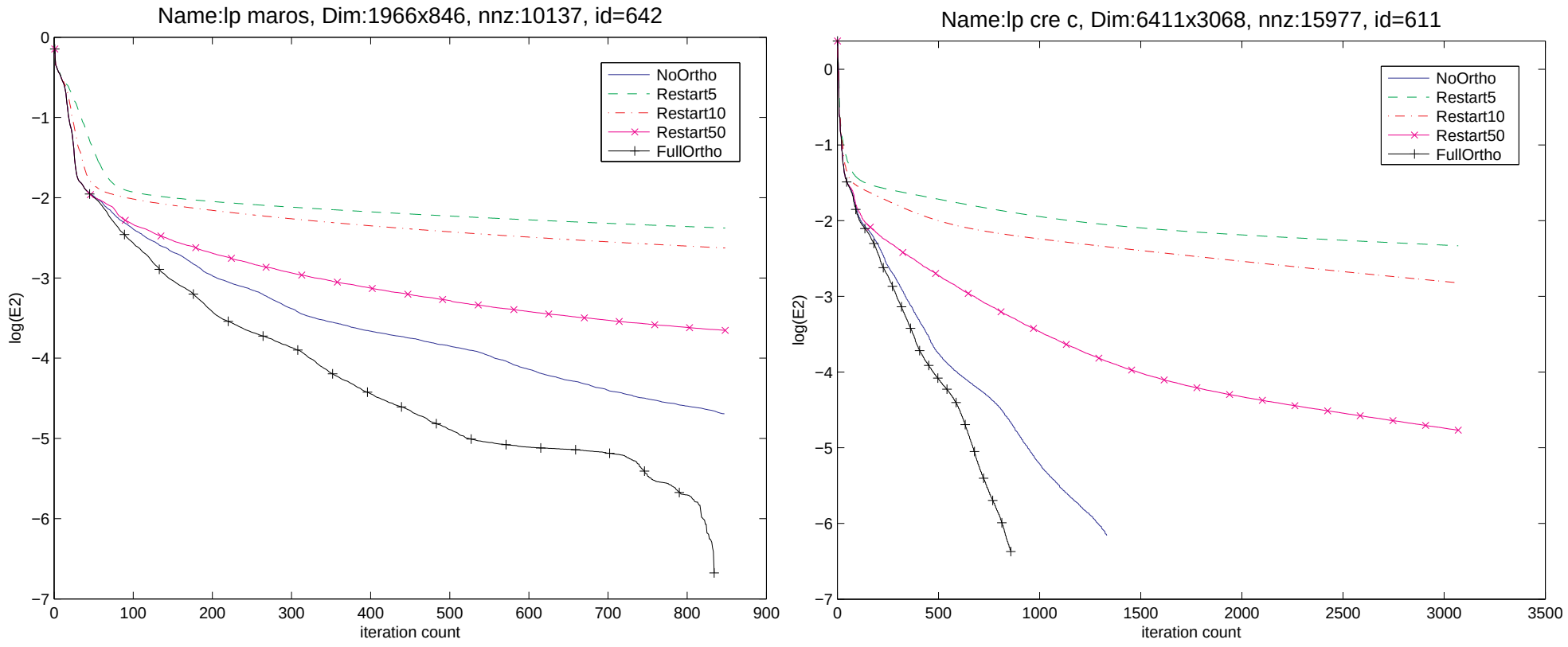

FIG. 7.10. *LSMR with reorthogonalized* $V_k$ *and restarting. Restart(*$\ell$*) with* $\ell = 5, 10, 50$ *is slower than standard LSMR with or without reorthogonalization. Problems* `lp_maros` *and* `lp_cre_c`*.*

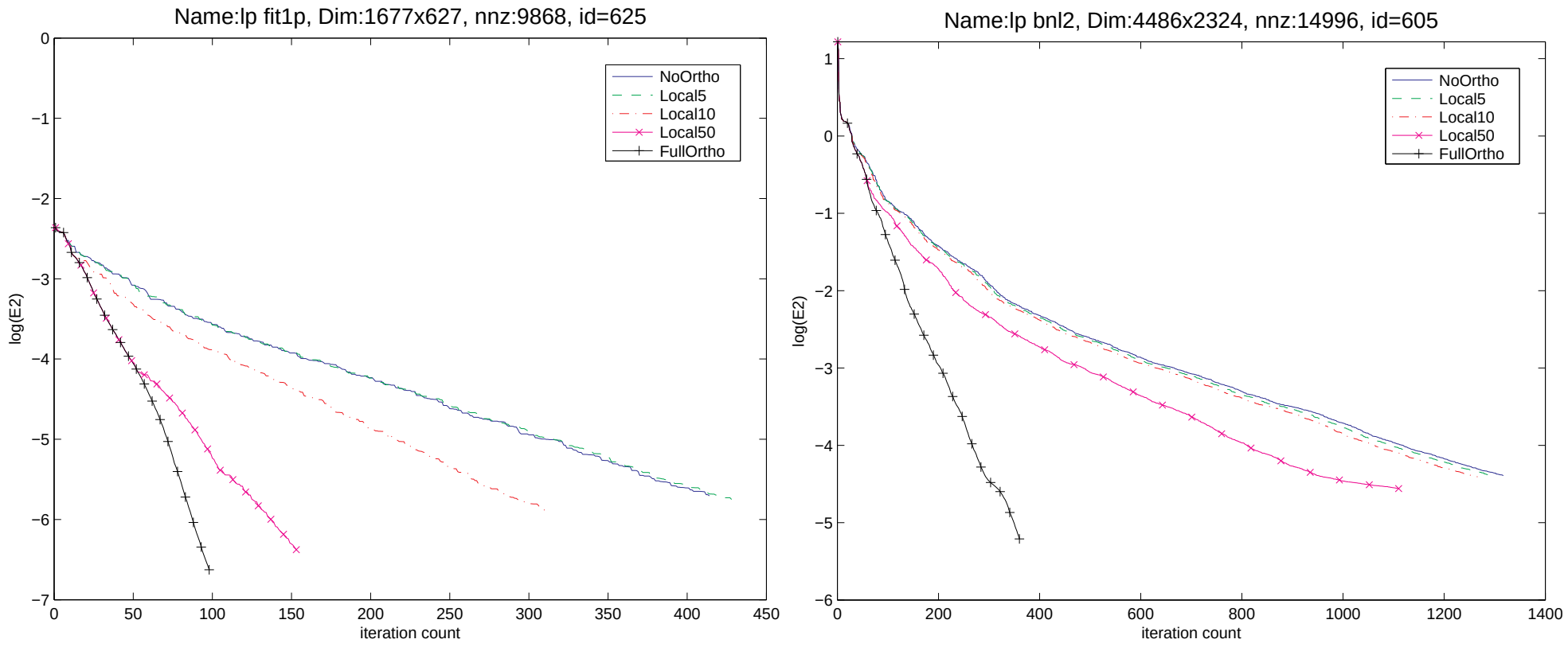

FIG. 7.11. *LSMR with local reorthogonalization of* $V_k$*. Local(*$\ell$*) with* $\ell = 5, 10, 50$ *illustrates reduced iterations as* $\ell$ *increases. Problems* `lp_fit1p` *and* `lp_bnl2`*.*

**7.3.1. Restarting.** To conserve storage, a simple approach is to restart the algorithm every $\ell$ steps, as with GMRES($\ell$) [20]. Figure 7.10 shows that restarting LSMR even with full reorthogonalization (of $V_k$) may lead to stagnation. In general, convergence with restarting is much slower than LSMR without reorthogonalization.

**7.3.2. Local reorthogonalization.** Here we reorthogonalize each new $v_k$ with respect to the previous $l$ vectors, where $l$ is a specified parameter. Figure 7.11 shows that $l = 5$ has little effect, but partial speedup was achieved with $l = 10$ and 50 in the two chosen cases. There is evidence of a useful storage-time tradeoff. The potential speedup depends strongly on the computational cost of $Av$ and $A^T u$.

**7.3.3. Partial reorthogonalization.** Larsen [19] uses partial reorthogonalization of both $V_k$ and $U_k$ within his PROPACK software for computing a set of singular values and vectors for a sparse rectangular matrix $A$. Similar techniques might prove helpful within LSMR. We leave this for future research.

**8. Summary.** We have presented LSMR, an iterative algorithm for square or rectangular systems, along with details of its implementation and experimental results to suggest that it has advantages over the widely adopted LSQR algorithm.

As in LSQR, theoretical and practical stopping criteria are provided for solving $Ax = b$ and $\min \|Ax - b\|$ with optional Tikhonov regularization, using estimates of $\|r_k\|$, $\|A^T r_k\|$, $\|x_k\|$, $\|A\|$ and $\text{cond}(A)$ that are cheaply computable. For LS problems, the Stewart backward error estimate $\|E_2\|$ (6.3) seems experimentally to be very close to the *optimal* backward error $\mu(x_k)$ at each LSMR iterate $x_k$ (section 6.2). This often allows LSMR to terminate significantly sooner than LSQR.

Experiments with full reorthogonalization have shown that the Golub-Kahan process retains high accuracy if the columns of either $V_k$ or $U_k$ are reorthogonalized. There is no need to reorthogonalize both. This discovery could be helpful for other uses of the Golub-Kahan process.

Matlab, Python, and Fortran 90 implementations of LSMR are available from [11]. They all allow local reorthogonalization of $V_k$.

**Acknowledgements.** We are grateful to Chris Paige for his helpful comments on reorthogonalization and other aspects of this work. We are also grateful to two referees for their extremely helpful and perceptive reviews. Further thanks go to Martin van Gijzen and Mike Botchev for their help with testing LSMR on square systems arising from convection-diffusion problems, to Sou-Cheng Choi for her helpful comments, and to Victor Pereyra for proposing that LSMR be used to terminate LSQR if a smaller final error $\|x - x_k\|$ is important.

**Appendix A. Proof of Lemma 3.1.**

The effects of the rotations $P_k$ and $\widetilde{P}_{k-1}$ can be summarized as

$$\widetilde{R}_k = \begin{pmatrix} \tilde{\rho}_1 & \tilde{\theta}_2 & & \\ & \ddots & \ddots & \\ & & \tilde{\rho}_{k-1} & \tilde{\theta}_k \\ & & & \dot{\rho}_k \end{pmatrix}, \quad \begin{array}{l} \begin{pmatrix} c_k & s_k \\ -s_k & c_k \end{pmatrix} \begin{pmatrix} \ddot{\beta}_k \\ 0 \end{pmatrix} = \begin{pmatrix} \hat{\beta}_k \\ \ddot{\beta}_{k+1} \end{pmatrix}, \\ \begin{pmatrix} \tilde{c}_k & \tilde{s}_k \\ -\tilde{s}_k & \tilde{c}_k \end{pmatrix} \begin{pmatrix} \dot{\rho}_{k-1} & & \dot{\beta}_{k-1} \\ \bar{\theta}_k & \bar{\rho}_k & \hat{\beta}_k \end{pmatrix} = \begin{pmatrix} \tilde{\rho}_{k-1} & \tilde{\theta}_k & \tilde{\beta}_{k-1} \\ 0 & \dot{\rho}_k & \dot{\beta}_k \end{pmatrix}, \end{array}$$

where $\ddot{\beta}_1 = \beta_1$, $\dot{\rho}_1 = \bar{\rho}_1$, $\dot{\beta}_1 = \hat{\beta}_1$ and where $c_k, s_k$ are defined in section 2.6.

We define $s^{(k)} = s_1 \dots s_k$ and $\bar{s}^{(k)} = \bar{s}_1 \dots \bar{s}_k$. Then from (3.3) and (2.4) we have $\widetilde{R}_k^T \tilde{t}_k = z_k = \begin{pmatrix} I_k & 0 \end{pmatrix} \bar{Q}_{k+1} e_{k+1} \bar{\beta}_1$. Expanding this and (3.1) gives

$$\widetilde{R}_k^T \tilde{t}_k = \begin{pmatrix} \bar{c}_1 \\ -\bar{s}_1 \bar{c}_2 \\ \vdots \\ (-1)^{k+1} \bar{s}^{(k-1)} \bar{c}_k \end{pmatrix} \bar{\beta}_1, \qquad \tilde{b}_k = \begin{pmatrix} \widetilde{Q}_k & \\ & 1 \end{pmatrix} \begin{pmatrix} c_1 \\ -s_1 c_2 \\ \vdots \\ (-1)^{k+1} s^{(k-1)} c_k \\ (-1)^{k+2} s^{(k)} \end{pmatrix} \beta_1,$$

and we see that

$$\tilde{\tau}_1 = \tilde{\rho}_1^{-1} \bar{c}_1 \bar{\beta}_1 \tag{A.1}$$

$$\tilde{\tau}_{k-1} = \tilde{\rho}_{k-1}^{-1} ((-1)^k \bar{s}^{(k-2)} \bar{c}_{k-1} \bar{\beta}_1 - \tilde{\theta}_{k-1} \tilde{\tau}_{k-2}) \tag{A.2}$$

$$\dot{\tau}_k = \dot{\rho}_k^{-1} ((-1)^{k+1} \bar{s}^{(k-1)} \bar{c}_k \bar{\beta}_1 - \tilde{\theta}_k \tilde{\tau}_{k-1}). \tag{A.3}$$

$$\dot{\beta}_1 = \hat{\beta}_1 = c_1 \beta_1 \tag{A.4}$$

$$\dot{\beta}_k = -\tilde{s}_{k-1} \dot{\beta}_{k-1} + \tilde{c}_{k-1} (-1)^{k-1} s^{(k-1)} c_k \beta_1 \tag{A.5}$$

$$\tilde{\beta}_k = \tilde{c}_k \dot{\beta}_k + \tilde{s}_k (-1)^k s^{(k)} c_{k+1} \beta_1. \tag{A.6}$$

We want to show by induction that $\tilde{\tau}_i = \tilde{\beta}_i$ for all $i$. When $i = 1$,

$$\tilde{\beta}_1 = \tilde{c}_1 c_1 \beta_1 - \tilde{s}_1 s_1 c_2 \beta_1 = \frac{\beta_1}{\tilde{\rho}_1}(c_1\bar{\rho}_1 - \bar{\theta}_2 s_1 c_2) = \frac{\beta_1}{\tilde{\rho}_1}\frac{\alpha_1}{\rho_1}\frac{\rho_1^2}{\bar{\rho}_1} = \frac{\bar{\beta}_1}{\tilde{\rho}_1}\frac{\rho_1}{\bar{\rho}_1} = \frac{\bar{\beta}_1}{\tilde{\rho}_1}\bar{c}_1 = \tilde{\tau}_1$$

where the third equality follows from the two lines below:

$$\begin{aligned} c_1\bar{\rho}_1 - \bar{\theta}_2 s_1 c_2 &= c_1\bar{\rho}_1 - \bar{\theta}_2 s_1 \frac{c_1\alpha_2}{\rho_2} = \bar{\rho}_1 - \bar{\theta}_2 s_1 \frac{\alpha_2}{\rho_2} = \frac{\alpha_1}{\rho_1}(\bar{\rho}_1 - \frac{1}{\rho_2}\bar{\theta}_2 s_1 \alpha_2) \\ \bar{\rho}_1 - \frac{1}{\rho_2}\bar{\theta}_2 s_1 \alpha_2 &= \bar{\rho}_1 - \frac{1}{\rho_2}(\bar{s}_1\rho_2)\theta_2 = \bar{\rho}_1 - \frac{\theta_2}{\bar{\rho}_1}\theta_2 = \frac{\bar{\rho}_1^2 - \theta_2^2}{\bar{\rho}_1} = \frac{\rho_1^2 + \theta_2^2 - \theta_2^2}{\bar{\rho}_1}. \end{aligned}$$

Suppose $\tilde{\tau}_{k-1} = \tilde{\beta}_{k-1}$. We consider the expression

$$\begin{aligned} s^{(k-1)} c_k \bar{\rho}_k^{-1} \bar{c}_{k-1}^2 \rho_k^2 \beta_1 &= \frac{\bar{c}_{k-1}\rho_k}{\bar{\rho}_k}(s^{(k-1)} c_k)\bar{c}_{k-1}\rho_k\beta_1 \\ &= \bar{c}_k \frac{\theta_2\cdots\theta_k\alpha_1}{\rho_1\cdots\rho_k}\frac{\rho_1\cdots\rho_{k-1}}{\bar{\rho}_1\cdots\bar{\rho}_{k-1}}\rho_k\beta_1 = \bar{c}_k\frac{\theta_2}{\bar{\rho}_1}\cdots\frac{\theta_k}{\bar{\rho}_{k-1}}\bar{\beta}_1 \\ &= \bar{c}_k \bar{s}_1\cdots\bar{s}_{k-1}\bar{\beta}_1 \qquad = \bar{c}_k \bar{s}^{(k-1)}\bar{\beta}_1. \end{aligned} \tag{A.7}$$

Applying the induction hypothesis on $\tilde{\tau}_k = \tilde{\rho}_k^{-1}\left((-1)^{k+1}\bar{s}^{(k-1)}\bar{c}_k\bar{\beta}_1 - \tilde{\theta}_k\tilde{\tau}_{k-1}\right)$ gives

$$\begin{aligned} \tilde{\tau}_k &= \tilde{\rho}_k^{-1}\left((-1)^{k+1}\bar{s}^{(k-1)}\bar{c}_k\bar{\beta}_1 - \tilde{\theta}_k\left(\tilde{c}_{k-1}\dot{\beta}_{k-1} + \tilde{s}_{k-1}(-1)^k s^{(k-1)} c_k\beta_1\right)\right) \\ &= \tilde{\rho}_k^{-1}\tilde{\theta}_k\tilde{c}_{k-1}\dot{\beta}_{k-1} + (-1)^{k+1}\tilde{\rho}_k^{-1}\left(\bar{s}^{(k-1)}\bar{c}_k\bar{\beta}_1 - \tilde{\theta}_k\tilde{s}_{k-1}s^{(k-1)}c_k\beta_1\right) \\ &= \tilde{\rho}_k^{-1}(\bar{\rho}_k\tilde{s}_{k-1})\tilde{c}_{k-1}\dot{\beta}_{k-1} + (-1)^{k+1}\tilde{\rho}_k^{-1}s^{(k-1)}\beta_1\left(\dot{\rho}_k\tilde{c}_{k-1}c_k - \bar{\theta}_{k+1}s_k c_{k+1}\right) \\ &= \tilde{c}_k\tilde{s}_{k-1}\dot{\beta}_{k-1} + (-1)^{k+1}s^{(k-1)}\beta_1\left(\tilde{c}_k\tilde{c}_{k-1}c_k - \tilde{s}_k s_k c_{k+1}\right) \\ &= \tilde{c}_k\left(-\tilde{s}_{k-1}\dot{\beta}_{k-1} + \tilde{c}_{k-1}(-1)^{k+1}s^{(k-1)}c_k\beta_1\right) + \tilde{s}_k(-1)^{k+1}s^{(k)}c_{k+1}\beta_1 \\ &= \tilde{c}_k\dot{\beta}_k + \tilde{s}_k(-1)^{k+1}s^{(k)}c_{k+1}\beta_1 = \tilde{\beta}_k \end{aligned}$$

with the second equality obtained by the induction hypothesis, and the fourth from

$$\begin{aligned} \bar{s}^{(k-1)}\bar{c}_k\bar{\beta}_1 - \tilde{\theta}_k\tilde{s}_{k-1}s^{(k-1)}c_k\beta_1 &= s^{(k-1)}c_k\bar{\rho}_k^{-1}\bar{c}_{k-1}^2\rho_k^2\beta_1 - (\tilde{s}_{k-1}\bar{\rho}_k)\tilde{s}_{k-1}s^{(k-1)}c_k\beta_1 \\ &= s^{(k-1)}\beta_1\frac{c_k}{\bar{\rho}_k}\left(\bar{c}_{k-1}^2\rho_k^2 - \tilde{s}_{k-1}^2\bar{\rho}_k^2\right) \\ &= s^{(k-1)}\beta_1\left(\dot{\rho}_k\tilde{c}_{k-1}c_k - \bar{\theta}_{k+1}s_k c_{k+1}\right), \end{aligned}$$

where the first equality follows from (A.7) and the last from

$$\begin{aligned} \bar{c}_{k-1}^2\rho_k^2 - \tilde{s}_{k-1}^2\bar{\rho}_k^2 &= \left(\bar{\rho}_k^2 - \theta_{k+1}^2\right) - \tilde{s}_{k-1}^2\bar{\rho}_k^2 = \bar{\rho}_k^2(1 - \tilde{s}_{k-1}^2) - \theta_{k+1}^2 = \bar{\rho}_k^2\tilde{c}_{k-1}^2 - \theta_{k+1}^2, \\ \frac{c_k}{\bar{\rho}_k}\bar{\rho}_k^2\tilde{c}_{k-1}^2 &= \bar{\rho}_k\tilde{c}_{k-1}^2 c_k = \dot{\rho}_k\tilde{c}_{k-1}c_k, \\ \frac{c_k}{\bar{\rho}_k}\theta_{k+1}^2 &= \frac{\theta_{k+1}}{\bar{\rho}_k}\theta_{k+1}c_k = \frac{\theta_{k+1}\rho_{k+1}}{\bar{\rho}_k}s_k\alpha_{k+1}\frac{c_k}{\rho_{k+1}} = \bar{\theta}_{k+1}s_k c_{k+1}. \end{aligned}$$

Therefore by induction, we know that $\tilde{\tau}_i = \tilde{\beta}_i$ for $i = 1, 2, \ldots$. From (3.3), we see that at iteration $k$, the first $k-1$ elements of $\tilde{b}_k$ and $\tilde{t}_k$ are equal. □